\documentclass[lettersize,journal]{IEEEtran}
\usepackage{amsmath,amsfonts}
\usepackage{algorithmic}
\usepackage{algorithm}
\usepackage{array}
\usepackage[caption=false,font=normalsize,labelfont=sf,textfont=sf]{subfig}
\usepackage{textcomp}
\usepackage{stfloats}
\usepackage{url}
\usepackage{verbatim}
\usepackage{graphicx}
\usepackage{cite}
\usepackage{array}
\usepackage{multirow}
\usepackage{booktabs}
\usepackage{color}
\usepackage{bbding}
\usepackage{epstopdf}
\usepackage[colorlinks,linkcolor=red]{hyperref}

\begin{document}

\title{Image Compressed Sensing Using\\ Non-local Neural Network}

\author{Wenxue~Cui,~\IEEEmembership{Student Member,~IEEE,}
        Shaohui~Liu,~\IEEEmembership{Member,~IEEE,}
        Feng~Jiang,~\IEEEmembership{Member,~IEEE,} \\
        and~Debin~Zhao,~\IEEEmembership{Member,~IEEE}


\thanks{This work was supported by the National Natural Science Foundation of China under Grant 61872116. ({\it{Corresponding author: Debin Zhao.}})}
\thanks{Wenxue Cui, Shaohui Liu, Feng Jiang and Debin Zhao are with the Department
of Computer Science and Technology, Harbin Institute of Technology, Harbin, 150001, China and also with the Peng Cheng Laboratory, Shenzhen, 518055, China (e-mail: wenxuecui@stu.hit.edu.cn; shliu@hit.edu.cn; fjiang@hit.edu.cn; dbzhao@hit.edu.cn).}}

\markboth{IEEE TRANSACTIONS ON MULTIMEDIA} 
{Shell \MakeLowercase{\textit{et al.}}: A Sample Article Using IEEEtran.cls for IEEE Journals}



\maketitle


\begin{abstract}
Deep network-based image Compressed Sensing (CS) has attracted much attention in recent years. However, the existing deep network-based CS schemes either reconstruct the target image in a block-by-block manner that leads to serious block artifacts or train the deep network as a black box that brings about limited insights of image prior knowledge. In this paper, a novel image CS framework using non-local neural network (NL-CSNet) is proposed, which utilizes the non-local self-similarity priors with deep network to improve the reconstruction quality. In the proposed NL-CSNet, two non-local subnetworks are constructed for utilizing the non-local self-similarity priors in the measurement domain and the multi-scale feature domain respectively. Specifically, in the subnetwork of measurement domain, the long-distance dependencies between the measurements of different image blocks are established for better initial reconstruction. Analogically, in the subnetwork of multi-scale feature domain, the affinities between the dense feature representations are explored in the multi-scale space for deep reconstruction. Furthermore, a novel loss function is developed to enhance the coupling between the non-local representations, which also enables an end-to-end training of NL-CSNet. Extensive experiments manifest that NL-CSNet outperforms existing state-of-the-art CS methods, while maintaining fast computational speed. The source code of plugging non-local module into certain CS network can be found: \url{https://github.com/WenxueCui/NL-CSNet-Pytorch}.
\end{abstract}

\begin{IEEEkeywords}
Image compressed sensing, non-local self-similarity prior, non-local neural network, convolutional neural networks (CNNs).
\end{IEEEkeywords}


\section{Introduction}

\IEEEPARstart{R}{ecent} years have seen significant interest in the compressed sensing (CS)~\cite{ref1, ref2}, which provides a new paradigm for signal acquisition that performs signal sampling and compression jointly. The CS theory implies that if a signal $x\in \mathbb{R}^{N}$ is sparse in a certain domain $\Psi$, it can be reconstructed with high probability from a small number of its linear measurements $y=\Phi x$, where $\Phi\in \mathbb{R}^{M\times N}$ is the sampling matrix with $M\ll N$ and $\frac{M}{N}$ is usually referred to as the sampling rate less than that determined by the Nyquist sampling theorem. The possible reduction of sampling rate is attractive for diverse practical applications, including but not limited to Magnetic Resonance Imaging (MRI)~\cite{ref3}, radar imaging~\cite{ref4} and sensor networks~\cite{ref5}.

In the study of CS, two main challenges are usually concerned: $1)$ the design of sampling matrix $\Phi$ for efficient signal acquisition and $2)$ the development of reconstruction solvers for recovering the original signal from its measurements~\cite{ref6}. In recent years, a great deal of algorithms have been proposed to deal with these two challenges. Specifically, for the sampling matrix, the Block-based CS (BCS)~\cite{ref7} is first proposed to squeeze the storage requirement of sampling matrix, in which the measurements are produced in a block-by-block sampling manner. Based on BCS, a variety of sampling matrices have been developed, such as the Gaussian Random Matrix (GRM)~\cite{ref8,ref9} and Local Structural-based Measurement Matrix (LSMM)~\cite{ref10}. However, both GRM and LSMM are signal independent, which may result in unsatisfactory reconstructed quality. For the CS reconstruction, many sparsity-regularized-based schemes have been presented, such as the greedy algorithms~\cite{ref11,ref12} and the convex-optimization schemes~\cite{ref13,ref14}. However, these CS reconstruction methods usually explore image priors to build model and then solve an optimization problem in an iterative fashion, which usually suffer from high computational complexity.

\begin{figure}[t]
\begin{center}
\includegraphics[width=1.7in]{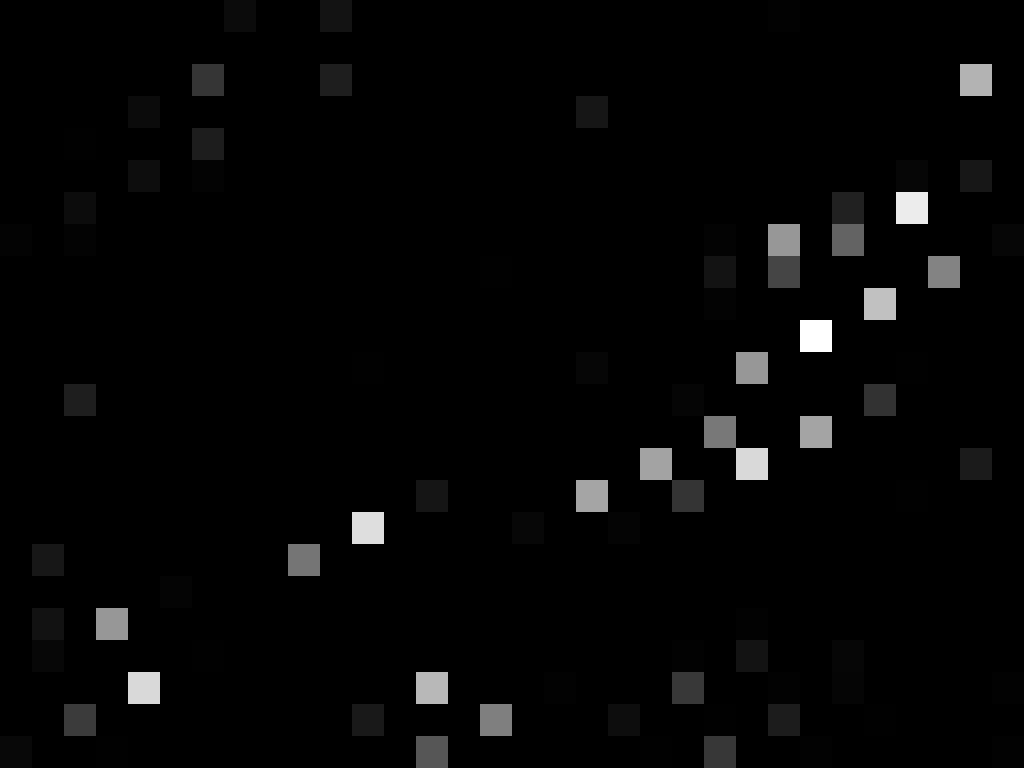}
\includegraphics[width=1.7in]{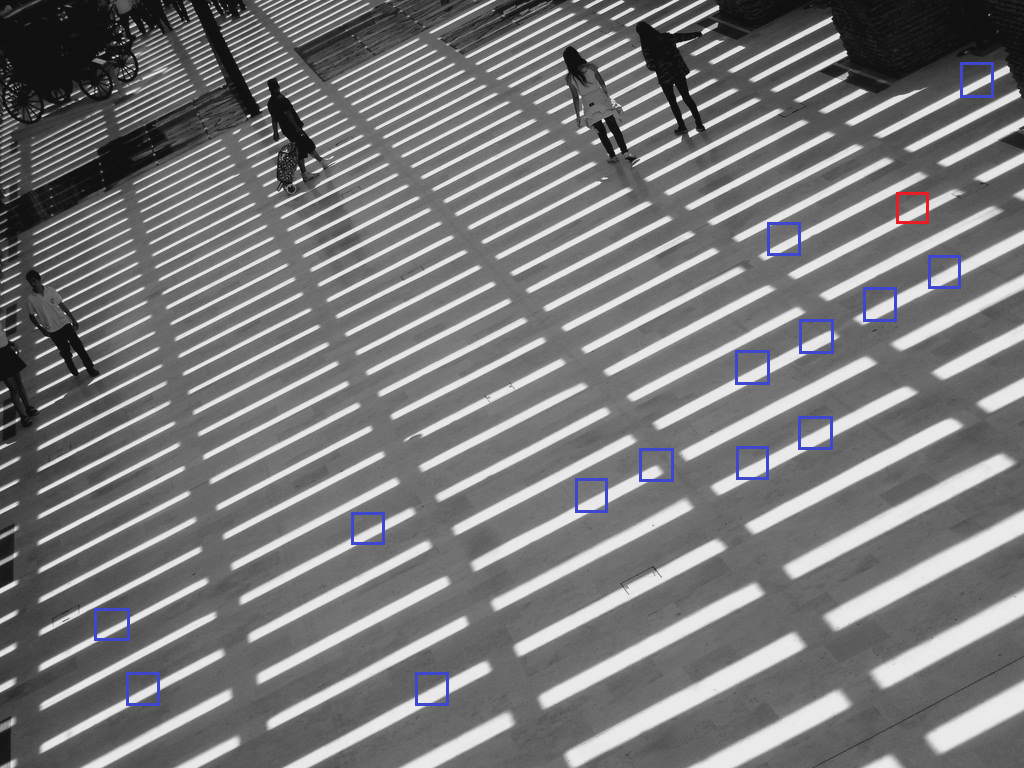}
\end{center}
\vspace{-0.09in} \scriptsize{\quad\quad\quad\quad\quad\quad\quad\quad\quad (a) \quad\quad\quad\quad\quad\quad\quad\quad\quad\quad\quad\quad\quad\quad\quad\quad (b)}
\vskip -0.1in
   \caption{The visualization of the learned affinity matrix in the non-local subnetwork of measurement domain (a) and its corresponding image patches on the original image (b). The red block is the current image patch and the blue blocks are the corresponding image patches mapped from the highlighted elements of the affinity matrix, which obviously have the similar structural textures.}
\vskip -0.24in
\label{fig:1}
\end{figure}

Recently, fueled by the powerful learning ability of deep neural networks, a series of deep network-based image CS methods have been proposed. For image sampling, the deep network-based sampling matrix can be optimized jointly with the reconstruction module during the training process~\cite{ref6}. For image reconstruction, the deep network-based schemes usually build a deep mapping from the measurement domain to the image domain, which can be roughly divided into two groups: \textbf{1) Training the deep network as a black box:} this kind of method trains the reconstruction network as a black box. Specifically, the early works~\cite{ref15,ref16,ref17,ref18} usually reconstruct the target image block-by-block and then splice the reconstructed image blocks together into a final image. However, the reconstruction block-by-block usually suffers from serious block artifacts especially at low sampling rates~\cite{ref6}. To relieve the block artifacts, some methods~\cite{ref6, ref19} attempt to concatenate all image blocks together in the initial reconstruction, and then complete a deep reconstruction in the global image space. Compared with the reconstruction block-by-block, these CS algorithms weaken the block artifacts and achieve higher reconstruction quality. However, these reconstruction networks generally establish a direct mapping from the measurement domain to the image domain in a rude manner and therefore result in limited insights of the image prior knowledge. \textbf{2) Interpretable CS reconstruction network:} this kind of method usually integrates the deep network with the iterative optimizers to enjoy a good interpretability. Motivated by the powerful learning capability of deep neural networks, some literatures~\cite{ref20,ref21,ref22,ref23} attempt to unfold the iterative optimization algorithms (e.g., iterative shrinkage-thresholding algorithm (ISTA)~\cite{ref24} and approximate message passing (AMP)~\cite{ref25}) onto networks to solve the CS reconstruction problem and achieve impressive performance. Obviously, by unfolding the optimization-based solvers, these deep unfolded methods have better interpretability, but these algorithms usually adopt a plain neural network architecture and therefore cannot fully exert the expressiveness of the proposed model for image reconstruction.

In this paper, a novel image CS framework using non-local neural network (dubbed NL-CSNet) is proposed, in which two non-local subnetworks are constructed for utilizing the non-local self-similarity priors in the measurement domain and the multi-scale feature domain respectively. Specifically, in the non-local subnetwork of measurement domain, we establish a long-distance reference between the measurements of different image blocks (as shown in Fig.~\ref{fig:1}), which efficiently explores the interblock correlations for better initial reconstruction. Analogically, in the non-local subnetwork of multi-scale feature domain, a long-range reference between the non-local structural textures is built, which explores the affinities between the dense feature representations in the multi-scale space for deep reconstruction. In addition, a novel loss function is proposed to enhance the coupling between the non-local representations, which also enables an end-to-end training of the proposed CS framework. Extensive experiments show that NL-CSNet outperforms existing state-of-the-art CS methods, while maintaining fast computational speed.


The main contributions are summarized as follows:

\textbf{1)} A novel image compressed sensing framework using non-local neural network is proposed, which utilizes the non-local self-similarity priors with deep network to improve the reconstruction quality.

\textbf{2)} In the measurement domain, the latent correlations between the measurements of different image blocks are explored for better initial reconstruction, which is favored for the following deep reconstruction.

\textbf{3)} In the multi-scale feature domain, the affinities between the dense feature representations are explored in the multi-scale space for deep reconstruction.

\textbf{4)} A novel loss function is designed to enhance the coupling between the non-local representations, which also provides an end-to-end training of the proposed CS framework.

The remainder of this paper is organized as follows: Section~\ref{section:a2} reviews the recent related works. Section~\ref{section:a3} elaborates the proposed CS framework, including the non-local neural network and the architectures of the networks embedding the non-local neural network in the measurement domain and the multi-scale feature domain respectively. Section~\ref{section:a4} provides the experimental results and Section~\ref{section:a6} concludes the paper. 

\begin{figure*}
\begin{center}
\includegraphics[width=6.9in]{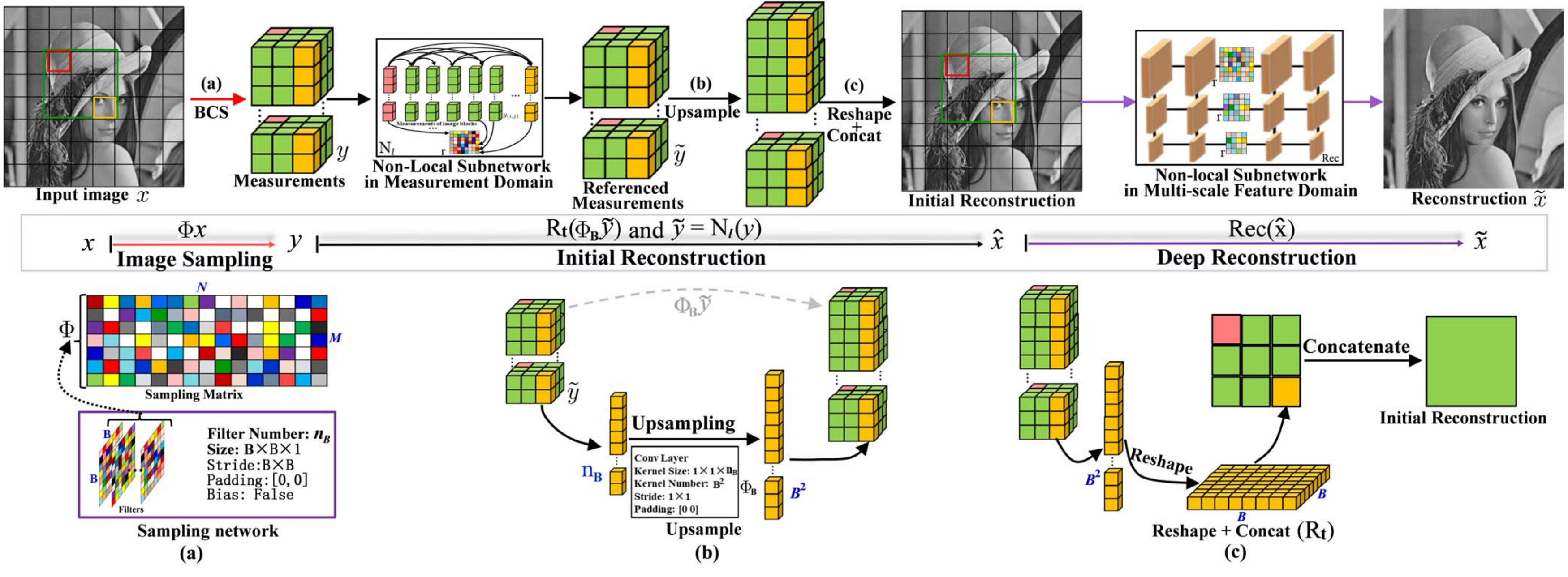}
\end{center}
\vskip -0.25in
   \caption{Diagram of our proposed image CS framework using non-local neural network. For CS reconstruction, two phases are included: initial reconstruction and deep reconstruction. The three sub-figures (a)-(c) show the details of three functional operators. The
blue text represents the dimensional description.}
\vskip -0.23in
\label{fig:2}
\end{figure*}


\vspace{-0.01in}
\section{Background and Related Works}

\label{section:a2}

\vspace{-0.01in}
\subsection{Image CS Reconstruction}
\vspace{-0.03in}
In CS task, the target image $x\in \mathbb{R}^{N}$ is reconstructed from its linear measurements $y\in \mathbb{R}^{M}$. Because $M$ $\ll$ $N$, this inverse problem is typically ill-posed. Recently, in order to solve this inverse problem, a great deal of CS methods have been proposed, which can be roughly grouped into two categories: optimization-based CS reconstruction methods and deep network-based CS reconstruction methods.

\textbf{Optimization-based CS Methods:} Given the linear measurements $y$, the traditional image CS reconstruction methods usually reconstruct the original image $x$ by solving an inverse optimization problem:
\vspace{-0.06in}
\begin{equation}
\tilde{x}=\mathop{\arg}\mathop{\min}_{x}\frac{1}{2} \| \Phi x-y \|_{2}^{2}+\lambda \| \Psi x \|_{\delta}
\label{Eq:1}
\vspace{-0.02in}
\end{equation}
where $\Psi x$ indicates the sparse coefficients with respect to the transform $\Psi$ and the sparsity is characterized by the $\delta$ norm. $\lambda$ is the regularization parameter to control the sparsity term. To solve Eq.~\eqref{Eq:1}, many sparsity-regularized based methods, such as the greedy algorithms~\cite{ref11,ref12} and the convex-optimization algorithms~\cite{ref13,ref14}, have been proposed. To further enhance the reconstructed quality, more sophisticated structures are established, including minimal total variation~\cite{ref26,ref27}, wavelet tree sparsity~\cite{ref28,ref29}, non-local image prior~\cite{ref30,ref31} and simple representations in adaptive bases~\cite{ref32}.
Many of these approaches have led to significant improvements. However, these optimization-based CS reconstruction algorithms usually suffer from high computational complexity because of their hundreds of iterations, thus limiting the practical applications of CS greatly.

\textbf{Deep Network-based CS Methods:} Driven by the powerful learning capability of deep neural networks, many deep network-based methods have been developed for image CS reconstruction, which can be roughly divided into two groups: 1) training the reconstruction network as a black box and 2) the interpretable CS reconstruction network.

For the first group of algorithms (training the reconstruction network as a black box), Mousavi \emph{et al.} first propose a stacked denoising autoencoder (SDA)~\cite{ref33} to capture statistical dependencies between the elements of the signal. However, the fully connected network (FCN) utilized in SDA leads to a huge number of learnable parameters. To relieve this problem, several Convolutional Neural Networks (CNNs) based reconstruction methods~\cite{ref15,ref16,ref18,ref34} are proposed, which usually build a direct mapping from the blocked measurements to the corresponding image blocks. However, these deep network-based CS algorithms usually bring about serious block artifacts~\cite{ref6,ref35} (especially at low sampling rates) because of their block-by-block reconstruction. In order to solve this problem, some works~\cite{ref16,ref17,ref18} try to append a de-blocking algorithm (e.g., BM3D~\cite{ref36}) after these block-by-block reconstruction methods, which usually introduce additional computational burden. To remove the block artifacts further, several literatures~\cite{ref6,ref19,ref35,ref37} attempt to explore the deep image priors in the whole image space. Specifically, these CS methods still adopt the block-by-block sampling, however during the reconstruction, they first concatenate all image blocks together in the initial reconstruction, and then complete a deep reconstruction in the whole image space. Recently, to enhance the applicability of CS framework, several scalable network architectures~\cite{ref19,ref38} are proposed, which achieve scalable sampling and reconstruction with only one model. More recently, a novel multi-channel deep network~\cite{ref39} for block-based image CS is proposed, in which the image blocks with different textural complexities are adaptively allocated different sampling rates. Compared to the block-by-block reconstruction methods, these CS algorithms usually achieve much higher reconstruction performance. However, the networks used in these methods are usually trained as a black box, which usually result in limited insights of the image prior knowledge.

For the second group of algorithms (the interpretable CS reconstruction network), the deep networks are usually integrated with the iterative optimizers to enjoy a better interpretability. For example, inspired by the denoising-based iterative thresholding (DIT) algorithm and the AMP algorithm, Metzler \emph{et al.}~\cite{ref20} first propose LDIT and LDAMP respectively for CS reconstruction. Subsequently, Zhang \emph{et al.}~\cite{ref21} propose a deep unfolded version of the popular algorithm ISTA~\cite{ref24}, dubbed ISTA-Net, which employs CNNs to learn the appropriate transformation operations and soft thresholding functions to reflect the sparsity of data. However, the performance of~\cite{ref21} is limited because of its random sampling manner and block-by-block reconstruction strategy. To enhance the reconstruction performance, the authors develop an enhanced version of~\cite{ref21}, dubbed OPINE-Net~\cite{ref23}, in which a learnable sampling matrix and an efficient deblocking strategy are adopted for boosting the reconstruction quality. More recently, by unfolding the iterative denoising process of the well-known AMP algorithm, Zhang \emph{et al.}~\cite{ref22} propose a deep unfolded model, dubbed AMP-Net, to solve the visual image CS problem. By unfolding the optimization-based iterative solvers onto networks, these deep unfolded methods have a better interpretability, but these algorithms usually adopt a plain network architecture and therefore cannot fully exert the expressiveness of the proposed model for image reconstruction.

\begin{figure}[b]
\vspace{-0.23in}
\begin{center}
\includegraphics[width=3.5in]{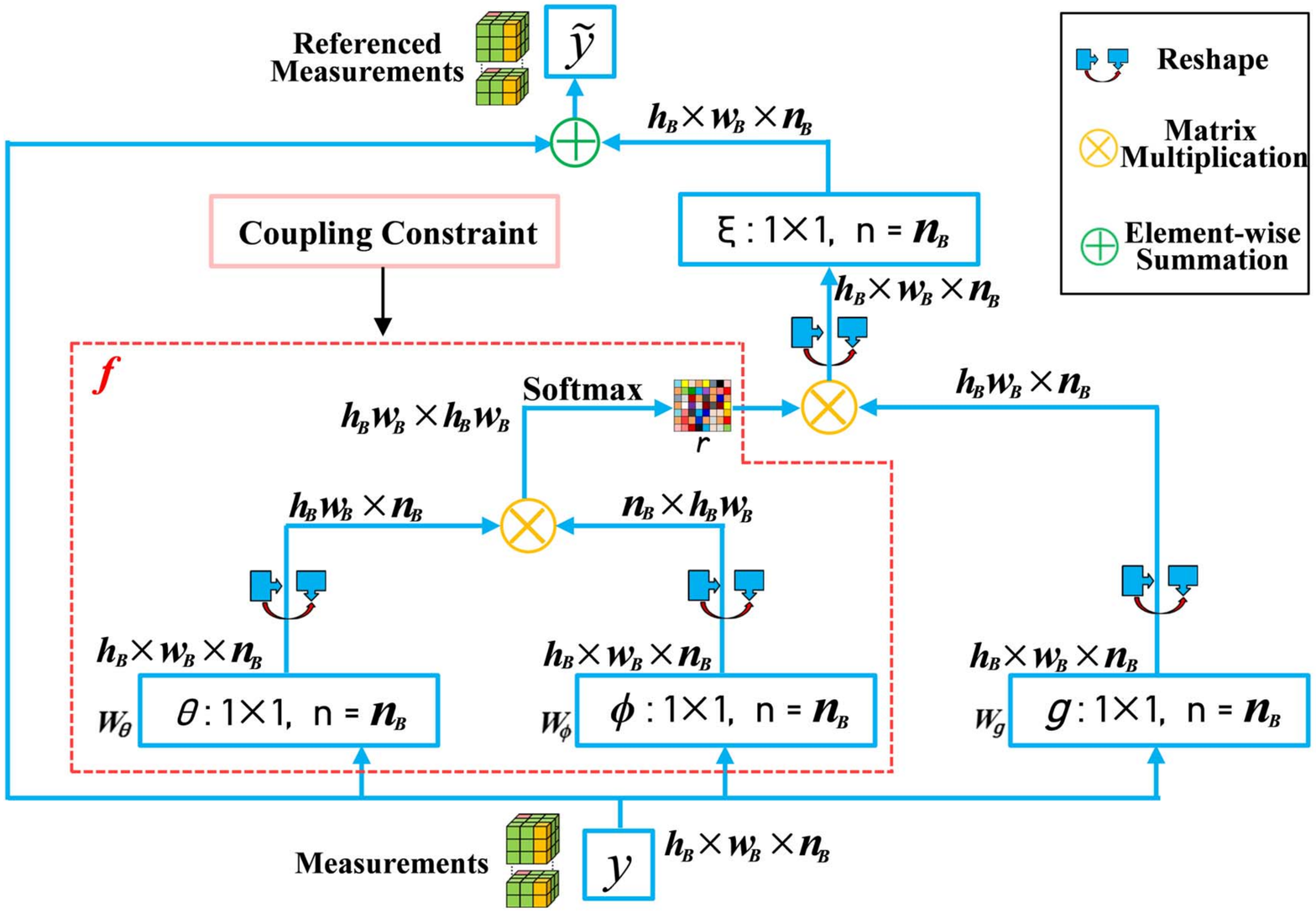}
\end{center}
\vskip -0.22in
   \caption{The details of non-local subnetwork in measurement domain. $n=n_{B}$ is the number of the convolutional filters of size $1\times 1$. The annotations indicate the dimensional information.}
\label{fig:3}
\end{figure}

\begin{figure*}
\begin{center}
\includegraphics[width=\textwidth]{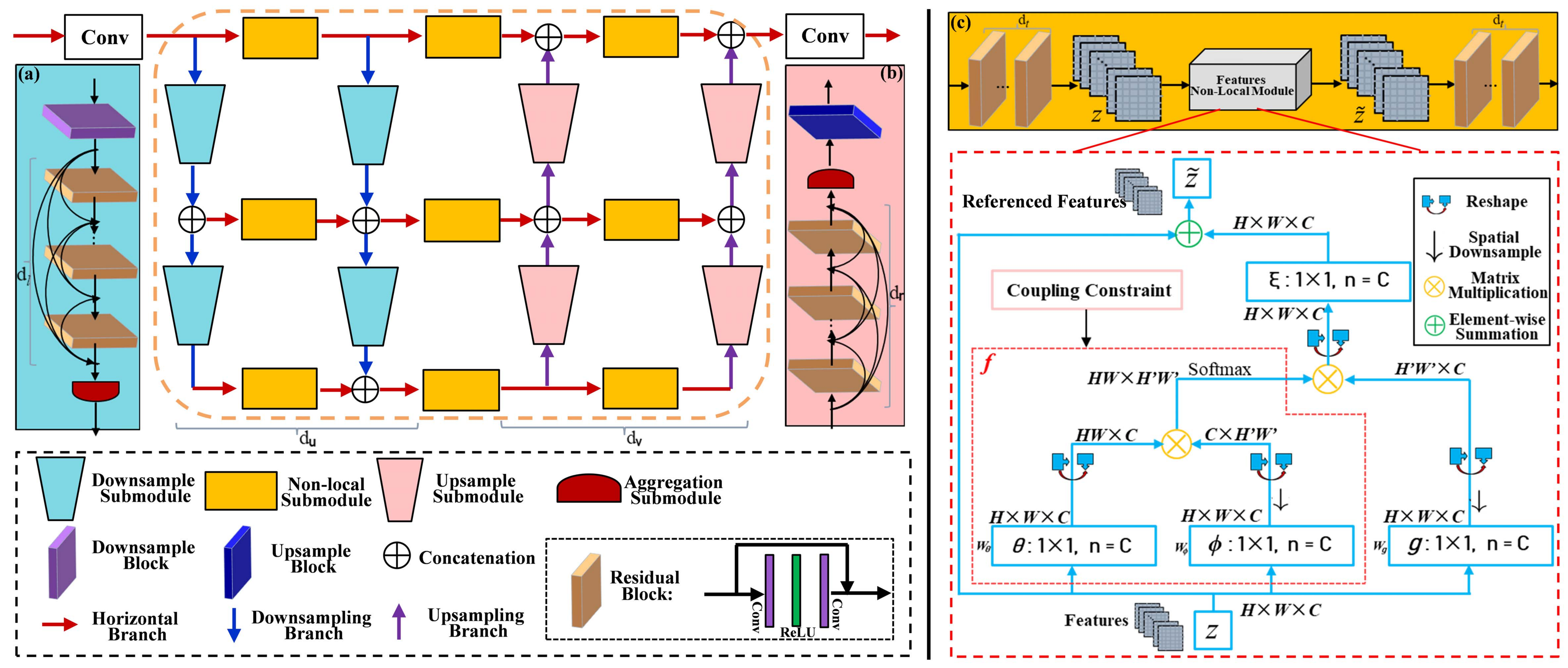}
\end{center}
\vskip -0.16in
   \caption{The architecture details of the proposed non-local subnetwork in multi-scale feature domain (MS-NLNet) in the deep reconstruction phase. (a)-(c) are the architecture details of three submodules: Downsample submodule, Upsample submodule and Non-local submodule.}
\vskip -0.2in
\label{fig:4}
\end{figure*}

Recently, motivated by the superiority of the CNN-based denoisers~\cite{ref40}, the plug-and-play algorithms~\cite{ref41,ref42,ref43,ref44} regularized by the deep denoising priors attract much attention for applying to diverse low-level vision tasks. The main idea is that, with the aid of variable splitting algorithms, such as half-quadratic splitting (HQS), it is possible to deal with the fidelity item and prior item separately~\cite{ref42}. Theoretically, the prior item only corresponds to a denoising subproblem~\cite{ref44}, which can be solved via a deep CNN denoiser. Based on the above statement, Romano \emph{et al.}~\cite{ref41} propose the paradigm of the Regularization by Denoising (RED): using the denoising engine (such as TNRD~\cite{ref45}) in defining the regularization of the inverse problem. In~\cite{ref42}, different CNN denoisers are trained to plug into Half Quadratic Splitting (HQS) algorithm for various low-level vision applications. In~\cite{ref43}, Liu \emph{et al.} propose to broaden the current denoiser-centric view of RED by considering priors corresponding to networks trained for more general artifact-removal. In~\cite{ref44}, Zhang \emph{et al.} expand their previous work~\cite{ref42} and propose a plug-and-play unrolled model using deep denoiser prior for image restoration. By integrating the deep neural network with the iterative optimizers, the deep unrolled models mentioned above can leverage the powerful expressiveness of CNNs. However, these methods usually adopt a plain convolutional neural network and therefore cannot fully characterize the rich image priors for image reconstruction.

\vspace{-0.12in}
\subsection{Non-local Self-Similarity Image Prior}
\vspace{-0.01in}

Recent studies show that the image prior models play an important role in diverse image processing tasks. The non-local self-similarity, as a well-known image prior, has been extensively studied, which depicts the repetitiveness of higher-level patterns (e.g., textures and structures) globally positioned in images. Inspired by the success of nonlocal means (NLM) denoising filter~\cite{ref46}, a series of nonlocal regularization terms for inverse problems exploiting nonlocal self-similarity property of natural images are emerging~\cite{ref47,ref48,ref49,ref50}. Due to the utilization of the non-local self-similarity priors, the methods with nonlocal regularization terms usually produce superior results, with sharper image edges and richer image details. Recently, the non-local self-similarity image prior is also applied in many optimization-based CS methods~\cite{ref30,ref31,ref51,ref52}. For example, Zhang \emph{et al.}~\cite{ref31} establish a novel sparse representation model of natural images by mining the nonlocal patches with the similar structures. Chen \emph{et al.}~\cite{ref52} exploit the non-local self-similarity patches and propose a low-rank based CS model for image reconstruction. By introducing the non-local self-similarity image prior, these non-local prior based methods establish the references between the non-local patches with the similar textures and obtain better CS reconstruction performance.

\begin{table*}[t]
\centering
\caption{Average PSNR and SSIM comparisons of different representative CS algorithms based on the random sampling matrix at various sampling rates on dataset Set5. Bold indicates the best result, and underline signifies the second best result.}
\label{tab:1}
\vspace{-0.12in}
\small

\begin{tabular}{p{3.25cm}<{\centering} | p{0.75cm}<{\centering} p{0.8cm}<{\centering} | p{0.75cm}<{\centering}  p{0.8cm}<{\centering} | p{0.75cm}<{\centering} p{0.8cm}<{\centering} | p{0.75cm}<{\centering} p{0.8cm}<{\centering} | p{0.75cm}<{\centering} p{0.8cm}<{\centering} | p{0.75cm}<{\centering} p{0.8cm}<{\centering}}
\toprule
\multirow{2}*{Algorithms} & \multicolumn{2}{c}{Rate=0.01} & \multicolumn{2}{c}{Rate=0.04} & \multicolumn{2}{c}{Rate=0.10} & \multicolumn{2}{c}{Rate=0.20} & \multicolumn{2}{c}{Rate=0.30} & \multicolumn{2}{c}{Avg.}\\
\cline{2-13}
&PSNR&SSIM&PSNR&SSIM&PSNR&SSIM&PSNR&SSIM&PSNR&SSIM&PSNR&SSIM\\
\midrule
\footnotesize{TV~\cite{ref53}}&15.53&0.4554&22.14&0.6076&27.07&0.7865&30.45&0.8709&32.75&0.9107&25.59&0.7262\\
\footnotesize{MH~\cite{ref32}}&18.08&0.4472&23.65&0.6337&28.57&0.8211&32.08&0.8881&34.06&0.9158&27.29&0.7412\\
\footnotesize{GSR~\cite{ref31}}&18.87&0.4909&24.80&0.7286&29.99&0.8654&\textbf{34.17}&\textbf{0.9257}&\textbf{36.83}&\underline{0.9492}&28.93&0.7920\\

\hline

\footnotesize{ReconNet}\tiny{$_{\rm \textcolor{red}{(CVPR2016)}}$}\footnotesize{~\cite{ref16}}&18.46&0.4492&23.54&0.6189&26.89&0.7518&29.55&0.8348&31.20&0.8738&25.93&0.7057\\
\footnotesize{I-Recon}\tiny{$_{\rm \textcolor{red}{(TCI2018)}}$}\footnotesize{~\cite{ref17}}&\underline{21.49}&\underline{0.5571}&\textbf{27.26}&0.7607&30.28&0.8496&33.12&0.9023&35.07&0.9357&\underline{29.44}&\underline{0.8011}\\
\footnotesize{ISTA-Net}\tiny{$_{\rm \textcolor{red}{(CVPR2018)}}$}\footnotesize{~\cite{ref21}}&18.48&0.4222&23.02&0.6428&28.53&0.8276&--\ --&--\ --&34.87&0.9354&--\ --&--\ --\\
\footnotesize{ISTA-Net$^{+}$}\tiny{$_{\rm \textcolor{red}{(CVPR2018)}}$}\footnotesize{~\cite{ref21}}&18.55&0.4408&23.45&0.6619&28.61&0.8315&--\ --&--\ --&35.45&0.9408&--\ --&--\ --\\
\footnotesize{NLR-CSNet}\tiny{$_{\rm \textcolor{red}{(TMM2020)}}$}\footnotesize{~\cite{ref54}}&21.00&--\ --&25.07&--\ --&29.23&--\ --&--\ --&--\ --&--\ --&--\ --&--\ --&--\ --\\
\footnotesize{DPA-Net}\tiny{$_{\rm \textcolor{red}{(TIP2020)}}$}\footnotesize{~\cite{ref37}}&--\ --&--\ --&26.63&\underline{0.7767}&\underline{30.32}&\underline{0.8713}&--\ --&--\ --&\underline{36.17}&\textbf{0.9495}&--\ --&--\ --\\

\footnotesize{DPIR}\tiny{$_{\rm \textcolor{red}{(TPAMI2021)}}$}\footnotesize{~\cite{ref44}}&17.68&0.4364&26.01&0.7565&29.97&0.8667&33.36&\underline{0.9206}&35.62&0.9475&28.53&0.7855\\

\midrule

NL-CSNet&\textbf{22.37}&\textbf{0.6031}&\underline{26.89}&\textbf{0.7786}&\textbf{30.60}&\textbf{0.8753}&\underline{33.41}&0.9202&35.74&0.9483&\textbf{29.80}&\textbf{0.8251}\\
\bottomrule

\end{tabular}

\vspace{-0.1in}
\label{tab:1}
\end{table*}

\begin{table*}[t]
\centering
\caption{Average PSNR and SSIM comparisons of different representative CS algorithms based on the learned sampling matrix at various sampling rates on dataset Set5. Bold indicates the best result, and underline signifies the second best result.}
\label{tab:2}
\vspace{-0.12in}
\small

\begin{tabular}{p{3.25cm}<{\centering} | p{0.75cm}<{\centering} p{0.8cm}<{\centering} | p{0.75cm}<{\centering}  p{0.8cm}<{\centering} | p{0.75cm}<{\centering} p{0.8cm}<{\centering} | p{0.75cm}<{\centering} p{0.8cm}<{\centering} | p{0.75cm}<{\centering} p{0.8cm}<{\centering} | p{0.75cm}<{\centering} p{0.8cm}<{\centering}}

\toprule

\multirow{2}*{Algorithms} & \multicolumn{2}{c}{Rate=0.01} & \multicolumn{2}{c}{Rate=0.04} & \multicolumn{2}{c}{Rate=0.10} & \multicolumn{2}{c}{Rate=0.20} & \multicolumn{2}{c}{Rate=0.30} & \multicolumn{2}{c}{Avg.}\\
\cline{2-13}
&PSNR&SSIM&PSNR&SSIM&PSNR&SSIM&PSNR&SSIM&PSNR&SSIM&PSNR&SSIM\\
\midrule

\footnotesize{CSNet}\tiny{$_{\rm \textcolor{red}{(ICME2017)}}$}\footnotesize{~\cite{ref55}}&24.02&0.6378&28.57&0.8226&32.30&0.9015&35.63&0.9451&37.90&0.9630&31.68&0.8540\\
\footnotesize{LapCSNet}\tiny{$_{\rm \textcolor{red}{(ICASSP2018)}}$}\footnotesize{~\cite{ref35}}&\underline{24.42}&\underline{0.6686}&--\ --&--\ --&32.44&0.9047&--\ --&--\ --&--\ --&--\ --&--\ --&--\ --\\
\footnotesize{SCSNet}\tiny{$_{\rm \textcolor{red}{(CVPR2019)}}$}\footnotesize{~\cite{ref19}}&24.21&0.6468&28.79&0.8314&32.77&0.9083&36.15&0.9487&38.45&0.9655&32.07&0.8601\\
\footnotesize{CSNet$^{+}$}\tiny{$_{\rm \textcolor{red}{(TIP2020)}}$}\footnotesize{~\cite{ref6}}&24.18&0.6478&28.64&0.8265&32.59&0.9062&36.05&0.9481&38.25&0.9644&31.94&0.8586\\
\footnotesize{BCS-Net}\tiny{$_{\rm \textcolor{red}{(TMM2020)}}$}\footnotesize{~\cite{ref39}}&22.98&0.6103&--\ --&--\ --&32.71&0.9030&36.12&0.9483&38.64&\underline{0.9694}&--\ --&--\ --\\
\footnotesize{OPINENet$^{+}$}\tiny{$_{\rm \textcolor{red}{(JSTSP2020)}}$}\footnotesize{~\cite{ref23}}&22.76&0.6194&\underline{29.03}&\underline{0.8440}&\underline{33.71}&\underline{0.9259}&--\ --&--\ --&--\ --&--\ --&--\ --&--\ --\\
\footnotesize{AMP-Net}\tiny{$_{\rm \textcolor{red}{(TIP2021)}}$}\footnotesize{~\cite{ref22}}&23.00&0.6488&28.85&0.8375&33.35&0.9162&\underline{36.64}&\underline{0.9532}&\textbf{39.07}&0.9671&\underline{32.18}&\underline{0.8646}\\

\midrule

NL-CSNet$^{\ast}$&\textbf{24.82}&\textbf{0.6771}&\textbf{29.61}&\textbf{0.8571}&\textbf{33.84}&\textbf{0.9312}&\textbf{36.91}&\textbf{0.9589}&\underline{38.86}&\textbf{0.9703}&\textbf{32.81}&\textbf{0.8789}\\
\bottomrule

\end{tabular}
\vspace{-0.2in}
\label{tab:2}
\end{table*}

By exploring the non-local self-similarity prior, several deep network-based CS schemes begin to utilize the non-local priors. For example, Li \emph{et al.}~\cite{ref34} propose a residual network with nonlocal constraint for image CS reconstruction, which considers the non-local self-similarity image prior and adds a non-local operation into the proposed network. While this CS network reconstructs the target image in a block-by-block manner from the measurements acquired by using the Gaussian random sampling matrix. Besides, the non-local operator in~\cite{ref34} only perceives the self-similar information inside the current image block, which ignores the correlations among different image blocks. Compared to~\cite{ref34}, the differences of the proposed framework are as follows: 1) Instead of block-by-block reconstruction manner, the proposed CS network reconstructs the target image from the measurements in the whole image space. 2) The sampling matrix of the proposed CS model is optimized jointly with the reconstruction network. 3) The proposed non-local module globally explores the self-similar knowledge in the entire image space, which significantly expands the receptive field. Recently, inspired by Deep Image Prior (DIP)~\cite{ref56}, Sun \emph{et al.}~\cite{ref54} propose a non-locally regularized CS network, in which the non-local prior and the deep network prior are both considered for image reconstruction. However, for each image, the network in~\cite{ref54} needs to be trained online driven by the two priors in an iterative fashion, which undoubtedly brings about a high computational cost and a lack of flexibility. By contrast, the proposed CS network generates final test model through pre-training. Specifically, in the training process, the proposed non-local module automatically characterizes the non-local image prior. In the testing process, the pre-trained model can directly capture the long-range dependencies in a feedforward fashion, while maintaining fast computational speed.

Transformer has demonstrated exemplary performance on a broad range of Natural Language Processing (NLP) related tasks~\cite{ref57,ref58}. Recently, the breakthroughs from Transformer networks in NLP domain has sparked great interest in the computer vision community~\cite{ref59}. However, these Transformers lack some of the inductive biases inherent to CNNs, such as translation equivariance and locality~\cite{ref59}, which makes their accuracy unsatisfactory on insufficient data sets. In addition, the inputs of Transformers are usually the linearly embedded patches, which accelerates the exploring of correlations between different image blocks, but weakens the learning ability of local representations to a certain extent. By contrast, the proposed network not only establishes the long-range references between different blocks in measurement domain, but also build the dependencies between dense representations in the multi-scale feature domain.

\begin{figure}[b]

\vspace{-0.22in}
\begin{minipage}[t]{0.110\textwidth}
\centering
\includegraphics[width=0.8in]{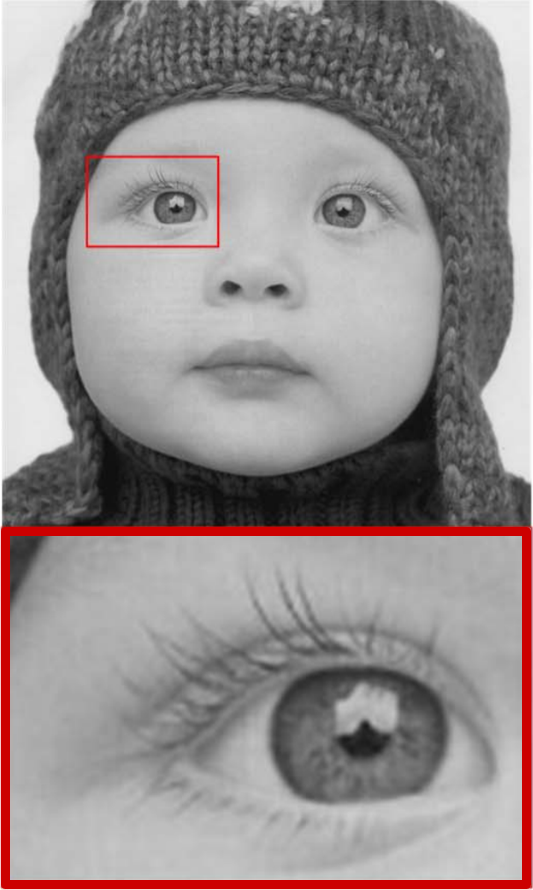}
\begin{scriptsize}
\centering
\vskip -0.38 cm \begin{tiny}Baby$\backslash$PSNR$\backslash$SSIM\end{tiny}

\end{scriptsize}
\end{minipage}
\hspace{-0.012in}
\begin{minipage}[t]{0.110\textwidth}
\centering
\includegraphics[width=0.8in]{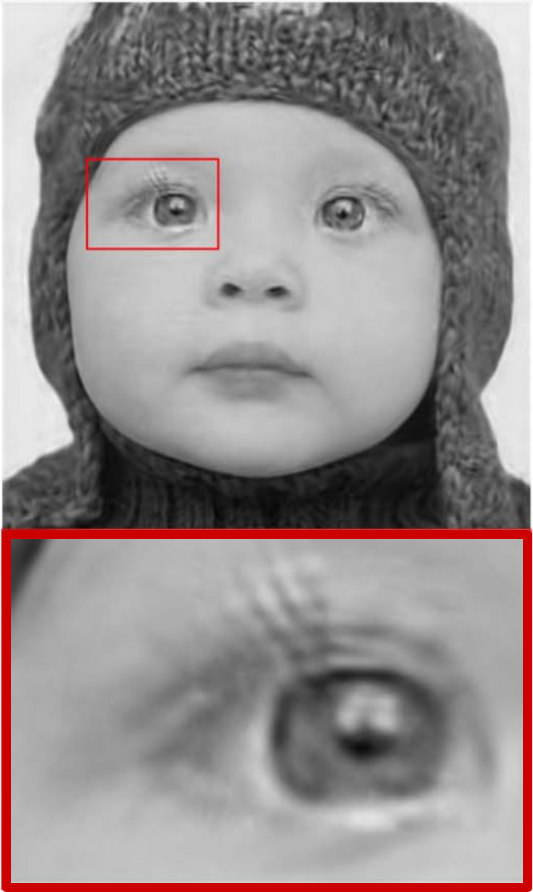}
\begin{scriptsize}
\centering
\vskip -0.52 cm \begin{tiny}MH~\cite{ref32}$\backslash$31.47$\backslash$0.8684\end{tiny}
\end{scriptsize}
\end{minipage}
\hspace{-0.012in}
\begin{minipage}[t]{0.110\textwidth}
\centering
\includegraphics[width=0.8in]{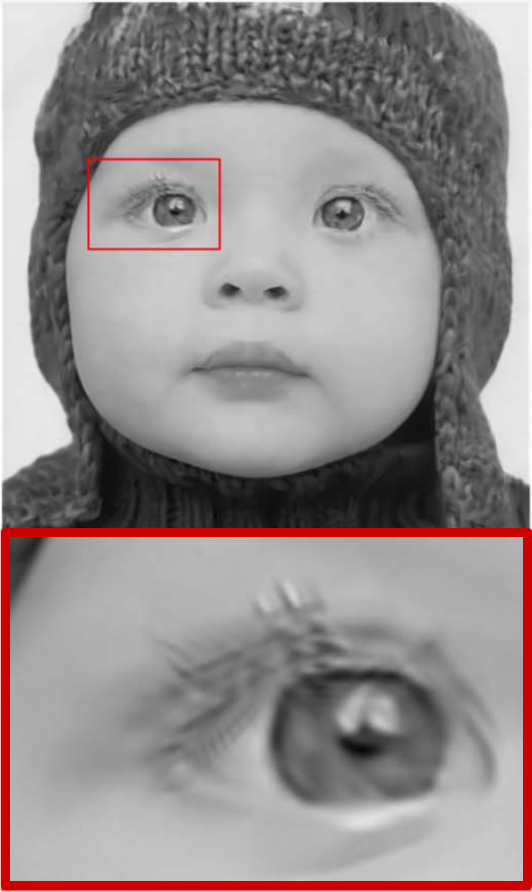}
\begin{scriptsize}
\centering
\vskip -0.52 cm \begin{tiny}GSR~\cite{ref31}$\backslash$32.18$\backslash$0.8832\end{tiny}
\end{scriptsize}
\end{minipage}
\hspace{-0.012in}
\begin{minipage}[t]{0.110\textwidth}
\centering
\includegraphics[width=0.8in]{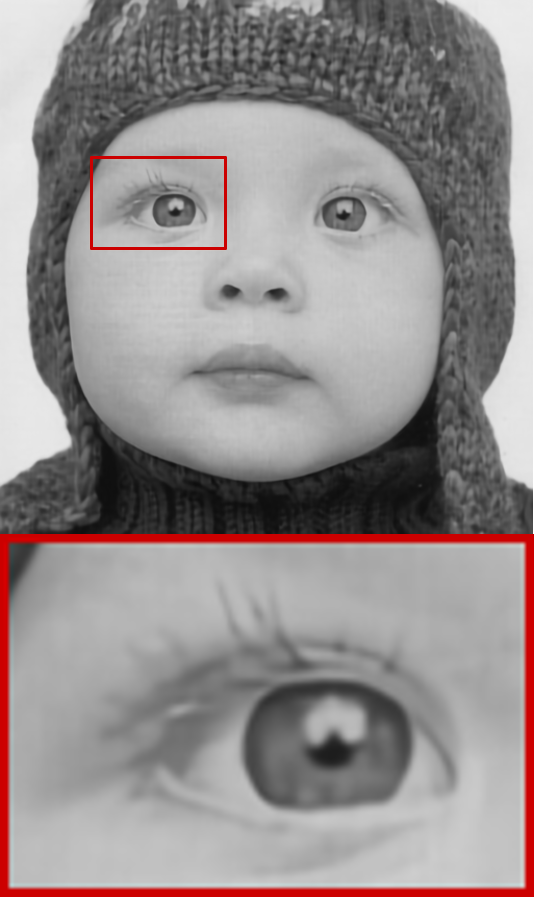}
\begin{scriptsize}

\centering
\vskip -0.38 cm \begin{tiny}NL-CSNet$\backslash$\textbf{33.09}$\backslash$\textbf{0.9025}\end{tiny}

\end{scriptsize}
\end{minipage}

\vspace{-0.01in} \caption{Visual comparisons of the proposed NL-CSNet and the other optimization-based CS methods on image \emph{Baby} from Set5~\cite{ref19} in case of sampling rate = 0.1.}
\label{fig:5}
\end{figure}

Graph convolutional network (GCN) is an effective strategy to establish long-range dependence for graph data. Recently, few works~\cite{ref60,ref61,ref62} apply GCN to image restoration tasks, and achieve impressive performance. However, the perceptive field in these GCN-based methods is usually limited within a predefined search window by considering the computational complexity~\cite{ref61}. Besides, a fixed number of similar references are explored in each window, which limits the flexibility and adaptability of the method. By contrast, the proposed non-local module perceives information in the whole image space. Furthermore, the proposed method separately explores coarse-grained references and find-grained dependencies in the measurement domain and the multi-scale feature domain, while the GCN-based methods usually capture non-local self-similar knowledge only at the dense feature space.

\begin{table*}[t]
\centering
\caption{Average PSNR and SSIM comparisons of different deep network-based CS algorithms using random sampling matrix at diverse sampling rates on dataset Set11. Bold indicates the best result, and underline signifies the second best result.}
\label{tab:3}
\vspace{-0.1in}

\small
\begin{tabular}{p{3.25cm}<{\centering} | p{0.75cm}<{\centering} p{0.8cm}<{\centering} | p{0.75cm}<{\centering}  p{0.8cm}<{\centering} | p{0.75cm}<{\centering} p{0.8cm}<{\centering} | p{0.75cm}<{\centering} p{0.8cm}<{\centering} | p{0.75cm}<{\centering} p{0.8cm}<{\centering} | p{0.75cm}<{\centering} p{0.8cm}<{\centering}}

\toprule
\multirow{2}*{Algorithms} & \multicolumn{2}{c}{Rate=0.01} & \multicolumn{2}{c}{Rate=0.04} & \multicolumn{2}{c}{Rate=0.10} & \multicolumn{2}{c}{Rate=0.20} & \multicolumn{2}{c}{Rate=0.30} & \multicolumn{2}{c}{Avg.}\\
\cline{2-13}
&PSNR&SSIM&PSNR&SSIM&PSNR&SSIM&PSNR&SSIM&PSNR&SSIM&PSNR&SSIM\\

\hline
\footnotesize{SDA}\tiny{$_{\rm \textcolor{red}{(ALLERTON2015)}}$}\footnotesize{~\cite{ref33}}&17.69&0.4376&21.05&0.5720&23.66&0.6794&--\ --&--\ --&24.77&0.7191&--\ --&--\ --\\
\footnotesize{ReconNet}\tiny{$_{\rm \textcolor{red}{(CVPR2016)}}$}\footnotesize{~\cite{ref16}}&17.54&0.4426&21.24&0.5748&24.07&0.6958&25.54&0.7719&28.72&0.8517&23.42&0.6674\\
\footnotesize{DR$^{2}$-Net}\tiny{$_{\rm \textcolor{red}{(CVPR2017)}}$}\footnotesize{~\cite{ref15}}&17.44&0.4294&20.80&0.5806&24.71&0.7175&--\ --&--\ --&30.52&--\ --&--\ --&--\ --\\
\footnotesize{IRCNN}\tiny{$_{\rm \textcolor{red}{(CVPR2017)}}$}\footnotesize{~\cite{ref42}}&7.70&--\ --&17.56&--\ --&24.02&--\ --&--\ --&--\ --&31.18&--\ --&--\ --&--\ --\\
\footnotesize{LDIT}\tiny{$_{\rm \textcolor{red}{(NIPS2017)}}$}\footnotesize{~\cite{ref20}}&17.58&0.4449&21.45&0.6075&25.56&0.7691&--\ --&--\ --&32.69&0.9223&--\ --&--\ --\\
\footnotesize{LDAMP}\tiny{$_{\rm \textcolor{red}{(NIPS2017)}}$}\footnotesize{~\cite{ref20}}&17.51&0.4409&21.30&0.5985&24.94&0.7483&--\ --&--\ --&32.01&0.9144&--\ --&--\ --\\
\footnotesize{I-Recon}\tiny{$_{\rm \textcolor{red}{(TCI2018)}}$}\footnotesize{~\cite{ref17}}&\underline{19.20}&\underline{0.5018}&\textbf{24.06}&0.7089&25.97&0.7888&\underline{27.92}&\underline{0.8457}&31.45&0.9135&25.72&0.7517\\
\footnotesize{DPDNN}\tiny{$_{\rm \textcolor{red}{(TPAMI2018)}}$}\footnotesize{~\cite{ref63}}&17.59&0.4459&21.11&0.6029&24.53&0.7392&--\ --&--\ --&32.06&0.9145&--\ --&--\ --\\
\footnotesize{ISTA-Net$^{+}$}\tiny{$_{\rm \textcolor{red}{(CVPR2018)}}$}\footnotesize{~\cite{ref21}}&17.45&0.4131&21.56&0.6240&26.49&0.8036&--\ --&--\ --&\textbf{33.70}&\underline{0.9382}&--\ --&--\ --\\
\footnotesize{NN}\tiny{$_{\rm \textcolor{red}{(TCI2020)}}$}\footnotesize{~\cite{ref64}}&17.67&0.4324&20.65&0.5525&22.99&0.6591&--\ --&--\ --&27.64&0.8095&--\ --&--\ --\\
\footnotesize{DPA-Net}\tiny{$_{\rm \textcolor{red}{(TIP2020)}}$}\footnotesize{~\cite{ref37}}&18.05&0.5011&23.50&\textbf{0.7205}&\underline{26.99}&\underline{0.8354}&--\ --&--\ --&--\ --&--\ --&--\ --&--\ --\\

\midrule

NL-CSNet&\textbf{19.59}&\textbf{0.5229}&\underline{23.74}&\underline{0.7097}&\textbf{27.24}&\textbf{0.8386}&\textbf{30.29}&\textbf{0.8910}&\underline{33.41}&\textbf{0.9386}&\textbf{26.85}&\textbf{0.7802}\\
\bottomrule

\end{tabular}
\vspace{-0.16in}
\label{tab:3}
\end{table*}


\vspace{-0.04in}
\section{Image Compressed Sensing Using Non-Local Neural Network}

\label{section:a3}

In this section, we first give an overview of the proposed whole CS framework, and then detail the structures of the non-local neural network. Finally, we introduce the architectures of the designed networks embedding the non-local neural network in the measurement domain and the multi-scale feature domain respectively.

\vspace{-0.08in}
\subsection{Overview of NL-CSNet}
\vspace{-0.01in}
Figure~\ref{fig:2} shows the whole network architecture of the proposed NL-CSNet. Specifically, for image sampling, a convolutional layer with specific parameter configurations (shown in Figure~\ref{fig:2}(a)) is utilized to imitate the block-based sampling process~\cite{ref6}, after which a series of measurements are generated. In order to reconstruct the target image from the measurements, a novel image CS reconstruction model using non-local neural network is proposed, in which two non-local subnetworks are constructed for exploring the non-local self-similarity priors in the measurement domain and the multi-scale feature domain respectively. More specifically, given the sampled measurements of all image blocks, the non-local subnetwork of measurement domain first establishes a long-distance reference between the measurements of different image blocks, which efficiently explores the interblock correlations in the measurement domain for better initial reconstruction. In fact, this kind of reference between the measurements of different image blocks is only a coarse non-local reference because of the non-overlapping block sampling of BCS. In order to further enhance the reconstructed quality from the initial reconstruction, the subnetwork of multi-scale feature domain is responsible for exploring the non-local self-similarity knowledge between the dense feature representations in the multi-scale space, which is capable of building a fine granular reference between the non-local structural textures for deep reconstruction. Furthermore, it is worth noting that a novel loss function is proposed to enhance the coupling of non-local information, which also enables an end-to-end training of the proposed CS framework.

\vspace{-0.06in}
\subsection{Non-Local Neural Network}
\vspace{-0.01in}

In the past few years, the non-local neural networks have been proposed in~\cite{ref65} to establish the long-distance references between the non-local representations with the similar structures. For example, given the current signal representation $x_{i}$ and according to the non-local mean operation~\cite{ref46}, the referenced information of $x_{i}$ by referencing the other signal representations can be expressed as:
\vspace{-0.04in}
\begin{equation}
\hat{x}_{i} = \frac{1}{\mathcal{C}(x)}\sum_{\forall j}f(x_{i}, x_{j})g(x_{j})
\label{Eq:2}
\vspace{-0.04in}
\end{equation}
where $f$ is a pairwise function to compute the affinities between the given representations $x_{i}$ and $x_{j}$. The unary function $g$ is used to compute a new representation of $x_{j}$. In fact, the non-local behavior in Eq.~\eqref{Eq:2} signifies that the all signal representations ($\forall$) are considered in the operation and the final response is normalized by a factor $\mathcal{C}(x)$.

\begin{table*}[t]
\centering
\caption{Average PSNR and SSIM comparisons of different deep network-based CS algorithms using learned sampling matrix at diverse sampling rates on dataset Set11. Bold indicates the best result, and underline signifies the second best result.}
\label{tab:4}
\vspace{-0.1in}
\small
\begin{tabular}{p{3.25cm}<{\centering} | p{0.75cm}<{\centering} p{0.8cm}<{\centering} | p{0.75cm}<{\centering}  p{0.8cm}<{\centering} | p{0.75cm}<{\centering} p{0.8cm}<{\centering} | p{0.75cm}<{\centering} p{0.8cm}<{\centering} | p{0.75cm}<{\centering} p{0.8cm}<{\centering} | p{0.75cm}<{\centering} p{0.8cm}<{\centering}}
\toprule
\multirow{2}*{Algorithms} & \multicolumn{2}{c}{Rate=0.01} & \multicolumn{2}{c}{Rate=0.04} & \multicolumn{2}{c}{Rate=0.10} & \multicolumn{2}{c}{Rate=0.20} & \multicolumn{2}{c}{Rate=0.30} & \multicolumn{2}{c}{Avg.}\\
\cline{2-13}
&PSNR&SSIM&PSNR&SSIM&PSNR&SSIM&PSNR&SSIM&PSNR&SSIM&PSNR&SSIM\\

\hline

\footnotesize{CSNet}\tiny{$_{\rm \textcolor{red}{(ICME2017)}}$}\footnotesize{~\cite{ref55}}&21.01&0.5560&25.23&0.7538&28.10&0.8514&31.36&0.9141&33.86&0.9448&27.91&0.8040\\
\footnotesize{SCSNet}\tiny{$_{\rm \textcolor{red}{(CVPR2019)}}$}\footnotesize{~\cite{ref19}}&\underline{21.04}&0.5562&25.48&0.7626&28.48&0.8616&31.82&0.9215&34.62&0.9511&28.29&0.8106\\
\footnotesize{CSNet$^{+}$}\tiny{$_{\rm \textcolor{red}{(TIP2020)}}$}\footnotesize{~\cite{ref6}}&21.03&\underline{0.5566}&25.41&0.7602&28.37&0.8580&31.66&0.9203&34.30&0.9490&28.15&0.8088\\
\footnotesize{BCS-Net}\tiny{$_{\rm \textcolor{red}{(TMM2020)}}$}\footnotesize{~\cite{ref39}}&20.88&0.5505&25.44&0.7425&29.43&0.8676&33.06&0.9283&35.06&0.9554&28.77&0.8089\\
\footnotesize{OPINENet$^{+}$}\tiny{$_{\rm \textcolor{red}{(JSTSP2020)}}$}\footnotesize{~\cite{ref23}}&20.02&0.5362&\underline{25.52}&\underline{0.7879}&\underline{29.81}&\underline{0.8904}&--\ --&--\ --&--\ --&--\ --&--\ --&--\ --\\
\footnotesize{AMP-Net$^{+}$}\tiny{$_{\rm \textcolor{red}{(TIP2021)}}$}\footnotesize{~\cite{ref22}}&20.20&0.5581&25.26&0.7722&29.40&0.8779&\underline{33.26}&\underline{0.9405}&\textbf{36.03}&\underline{0.9586}&\underline{28.83}&\underline{0.8215}\\

\midrule
NL-CSNet$^{\ast}$&\textbf{21.96}&\textbf{0.6005}&\textbf{26.26}&\textbf{0.8108}&\textbf{30.05}&\textbf{0.8995}&\textbf{33.52}&\textbf{0.9440}&\underline{35.68}&\textbf{0.9606}&\textbf{29.49}&\textbf{0.8431}\\
\bottomrule

\end{tabular}
\vspace{-0.08in}
\label{tab:4}
\end{table*}

\begin{table*}[t]

\centering
\caption{Average PSNR and SSIM comparisons of recent deep network-based CS algorithms using learned sampling matrix at various sampling rates on dataset Set14. Bold indicates the best result, and underline signifies the second best result.}
\label{tab:5}
\vspace{-0.1in}
\small

\begin{tabular}{p{0.78cm}<{\centering} | p{0.67cm}<{\centering} | p{0.85cm}<{\centering} p{0.92cm}<{\centering} | p{0.85cm}<{\centering}  p{0.92cm}<{\centering} | p{0.85cm}<{\centering} p{0.92cm}<{\centering} | p{0.85cm}<{\centering} p{0.92cm}<{\centering} | p{0.85cm}<{\centering} p{0.92cm}<{\centering} | p{0.85cm}<{\centering} p{0.92cm}<{\centering}}

\toprule
\multirow{2}*{Data} & \multirow{2}*{Rate} & \multicolumn{2}{c}{CSNet~\cite{ref55}} & \multicolumn{2}{c}{SCSNet~\cite{ref19}} & \multicolumn{2}{c}{CSNet$^{+}$~\cite{ref6}} & \multicolumn{2}{c}{OPINE-Net$^{+}$~\cite{ref23}} & \multicolumn{2}{c}{AMP-Net~\cite{ref22}} & \multicolumn{2}{c}{NL-CSNet$^{\ast}$}\\
\cline{3-14}
&&PSNR&SSIM&PSNR&SSIM&PSNR&SSIM&PSNR&SSIM&PSNR&SSIM&PSNR&SSIM\\
\midrule

\multirow{5}*{Set14}&0.01&22.79&0.5628&\underline{22.87}&0.5631&22.83&0.5630&22.30&0.5508&22.60&\underline{0.5723}&\textbf{23.61}&\textbf{0.5862}\\

&0.04&26.05&0.7164&26.24&0.7210&26.11&0.7196&\underline{26.67}&\underline{0.7306}&26.60&0.7212&\textbf{27.11}&\textbf{0.7460}\\
&0.10&28.91&0.8119&29.22&0.8181&29.13&0.8169&\underline{29.94}&\underline{0.8415}&29.87&0.8130&\textbf{30.16}&\textbf{0.8527}\\

&0.20&31.86&0.8908&32.19&0.8945&32.15&0.8941&--\ --&--\ --&\underline{32.72}&\underline{0.9024}&\textbf{32.96}&\textbf{0.9150}\\
&0.30&34.00&0.9276&34.51&0.9311&34.34&0.9297&--\ --&--\ --&\textbf{35.23}&\underline{0.9364}&\underline{34.88}&\textbf{0.9405}\\

\midrule

Avg.&---&28.72&0.7819&29.01&0.7856&28.91&0.7847&--\ --&--\ --&\underline{29.40}&\underline{0.7891}&\textbf{29.74}&\textbf{0.8081}\\
\bottomrule

\end{tabular}
\vspace{-0.18in}
\label{tab:5}
\end{table*}

\begin{figure}[b]
\vskip -0.16in
\begin{center}

\includegraphics[width=0.835in]{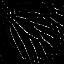}
\includegraphics[width=0.835in]{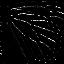}
\includegraphics[width=0.835in]{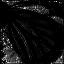}
\includegraphics[width=0.835in]{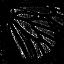}
\\
\vspace{0.02in}
\includegraphics[width=0.835in]{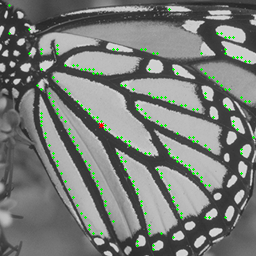}
\includegraphics[width=0.835in]{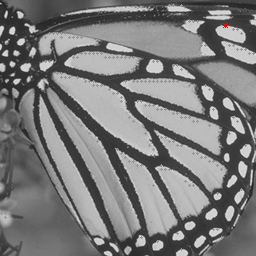}
\includegraphics[width=0.835in]{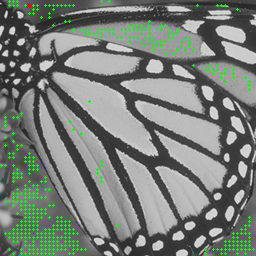}
\includegraphics[width=0.835in]{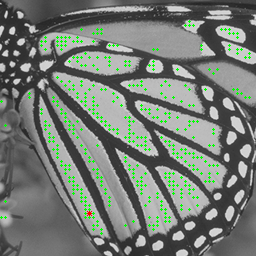}

\vskip -0.02in \tiny{\quad (a) \quad\quad\quad\quad\quad\quad\quad\quad\quad\quad\quad(b) \quad\quad\quad\quad\ \quad\quad\quad\quad\quad\quad\quad (c) \quad\quad\quad\quad\quad\quad\quad\quad\quad\quad \quad (d)}

\end{center}
\vskip -0.12in
   \caption{The visualization of the learned affinity matrix in the non-local subnetwork of multi-scale feature domain (top) and its corresponding positions on the original image (bottom). The red points are the current locations and the green points are the corresponding positions mapped from the highlighted elements of affinity matrix, around which the similar texture is maintained. (a), (b) are edge areas, and (c), (d) are smooth areas.}
\label{fig:6}
\end{figure}

Obviously, $f(x_{i}, x_{j})$ and $f(x_{j}, x_{i})$ are a pair of symmetrical affinities and they are both actually closely related to the similarity between the given two representations $x_{i}$ and $x_{j}$. In most non-local prior based iterative CS literatures~\cite{ref31}, Euclidean distance is usually selected as the similarity criterion between different image patches, which actually is a directionless scalar metric. In other words, given two signal representations $x_{i}$ and $x_{j}$ with the similar structures, there actually is a correspondence between $f(x_{i}, x_{j})$ and $f(x_{j}, x_{i})$. For example, if $x_{i}$ is similar to $x_{j}$, we can obtain that $x_{j}$ is also similar to $x_{i}$. Therefore, there is a coupling relationship between $f(x_{i}, x_{j})$ and $f(x_{j}, x_{i})$. Unfortunately, in existing non-local neural networks~\cite{ref65} or their counterpart variants~\cite{ref66,ref67}, the affinity is generally computed directly in the embedding space (mapped through two linear matrices $\theta$ and $\phi$), such as the embedded gaussian or the embedded dot product, which does not consider the coupling between $f(x_{i}, x_{j})$ and $f(x_{j}, x_{i})$. In our non-local module, we propose a new constraint as follows:
\vspace{-0.05in}
\begin{equation}
\hat{f}(x_{i}, x_{j})\hspace{-0.03in}=\hspace{-0.03in}\mathop{\arg\min}\limits_{f} \hspace{-0.03in}\|f(x_{i}, x_{j})\hspace{-0.02in}-\hspace{-0.02in}f(x_{j}, x_{i})\|_{l}
\label{eq:4}
\vspace{-0.06in}
\end{equation}
where the coupling (i.e., the distance) between $f(x_{i}, x_{j})$ and $f(x_{j}, x_{i})$ is characterized by the $l$ norm. Through the above constraint, the consistency between $f(x_{i}, x_{j})$ and $f(x_{j}, x_{i})$ is maintained to a certain extent, which facilitates a mutual reference between the non-local information. It is worth noting that the constraint in Eq.~\eqref{eq:4} can be embedded into the loss function of the proposed NL-CSNet, which will be elaborated in the Subsection IV-A. The following two subsections introduce the architectures of networks embedding the non-local neural network in the measurement domain and the multi-scale feature domain respectively.

\begin{table}[b]

\vspace{-0.24in}
\centering
\caption{Average running time (in seconds) of different CS algorithms
for reconstructing a 256 $\times$ 256 image.}
\label{tab:6}
\vspace{-0.1in}
\begin{tabular}{p{1.78cm}<{\centering} | p{1.25cm}<{\centering}  p{1.2cm}<{\centering} | p{1.25cm}<{\centering}  p{1.2cm}<{\centering}}
\toprule

\multirow{2}*{\footnotesize{Algorithm}} & \multicolumn{2}{c|}{\footnotesize{Rate=0.01}} & \multicolumn{2}{c}{\footnotesize{Rate=0.1}}\\
\cline{2-5}

&\footnotesize{CPU}&\footnotesize{GPU}&\footnotesize{CPU}&\footnotesize{GPU}\\
\midrule

\footnotesize{TV~\cite{ref53}}&\small{2.4006}&\small{---}&\small{2.7405}&\small{---}\\

\footnotesize{MH~\cite{ref32}}&\small{23.1006}&\small{---}&\small{19.0405}&\small{---}\\

\footnotesize{GSR~\cite{ref31}}&\small{235.6297}&\small{---}&\small{230.4755}&\small{---}\\
\midrule

\footnotesize{SDA~\cite{ref33}}&\small{---}&\small{0.0045}&\small{---}&\small{0.0029}\\

\footnotesize{ReconNet~\cite{ref16}}&\small{0.5193}&\small{0.0244}&\small{0.5258}&\small{0.0289}\\

\footnotesize{ISTA-Net\footnotesize{~\cite{ref21}}}&\small{0.9230}&\small{0.0390}&\small{0.9240}&\small{0.0395}\\
\footnotesize{\hspace{-0.05in}ISTA-Net$^{+}$\footnotesize{~\cite{ref21}}}&\small{1.3750}&\small{0.0470}&\small{1.3820}&\small{0.0485}\\
\footnotesize{CSNet~\cite{ref55}}&\small{0.2950}&\small{0.0157}&\small{0.3014}&\small{0.0168}\\
\footnotesize{SCSNet~\cite{ref19}}&\small{0.5103}&\small{0.1050}&\small{0.5146}&\small{0.1332}\\
\footnotesize{CSNet$^{+}$~\cite{ref6}}&\small{0.8960}&\small{0.0262}&\small{0.9024}&\small{0.0287}\\
\midrule
\footnotesize{NL-CSNet}&\small{0.8246}&\small{0.0899}&\small{0.9075}&\small{0.0964}\\

\bottomrule

\end{tabular}
\vspace{-0.01in}
\label{tab:6}
\end{table}

\vspace{-0.04in}
\subsection{Non-Local Subnetwork in Measurement Domain}

In BCS, the image $x$ with size $w \times h$ is first divided into non-overlapping blocks $x_{(i,j)}$ of size $B\times B$, where $i\in \{1,2,...,w_{B}\}$ ($w_{B}=\frac{w}{B}$) and $j\in \{1,2,...,h_{B}\}$ ($h_{B}=\frac{h}{B}$) are the position indexes of the current image block. Then, a sampling matrix $\Phi$ of size $n_{B}\times B^{2}$ is usually used to acquire the CS measurements, i.e., $y_{(i,j)}=\Phi x_{(i,j)}$, where $n_{B}=\lfloor\frac{M}{N}B^{2}\rfloor$ and $\frac{M}{N}$ is the given sampling rate. In fact, each row of the sampling matrix $\Phi$ can be considered as a filter. Therefore, we can use a series of convolutional operations to mimic the sampling process~\cite{ref6}. Specifically, in our CS framework, a convolutional layer with the specific configurations as shown in Fig.~\ref{fig:2}(a) is utilized to imitate the sampling process. Besides, it is worth noting that the measurements between different image blocks are actually independent because of the block-by-block sampling manner in BCS. In order to establish the correlations between the measurements of different image blocks, we embed the non-local neural network into the measurement domain, i.e.,


\begin{figure*}[t]
\begin{center}
\includegraphics[width=1.385in]{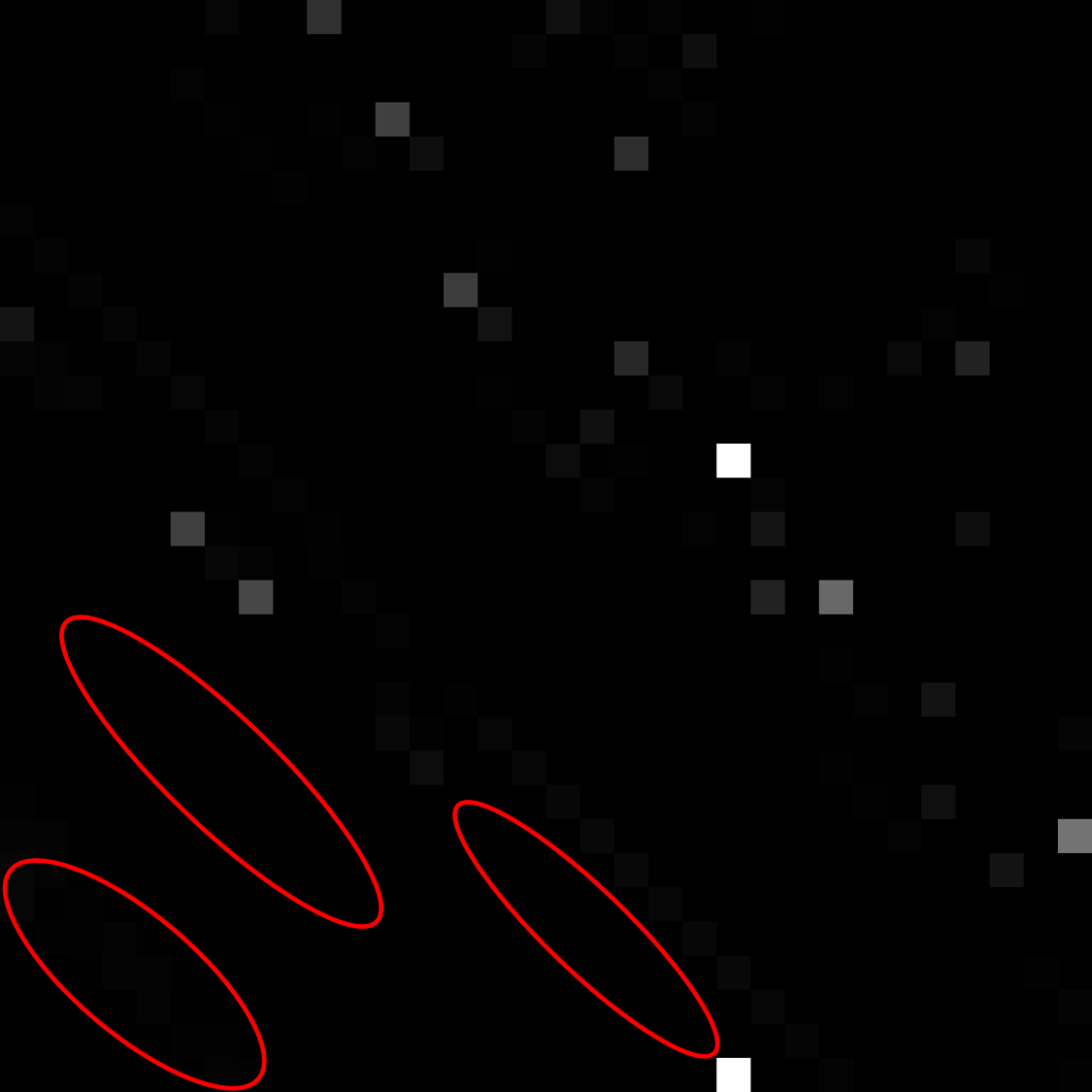}
\includegraphics[width=1.385in]{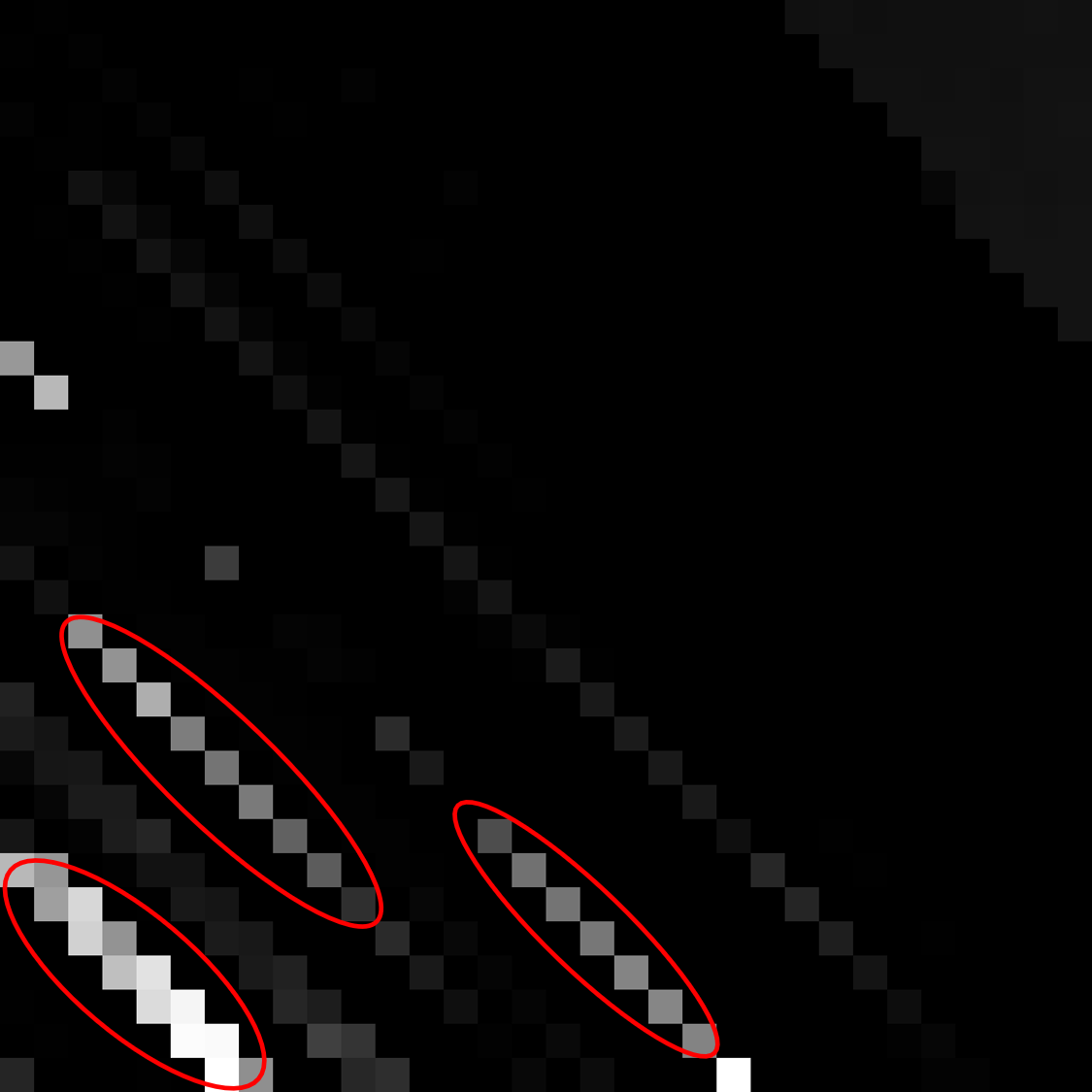}
\includegraphics[width=1.385in]{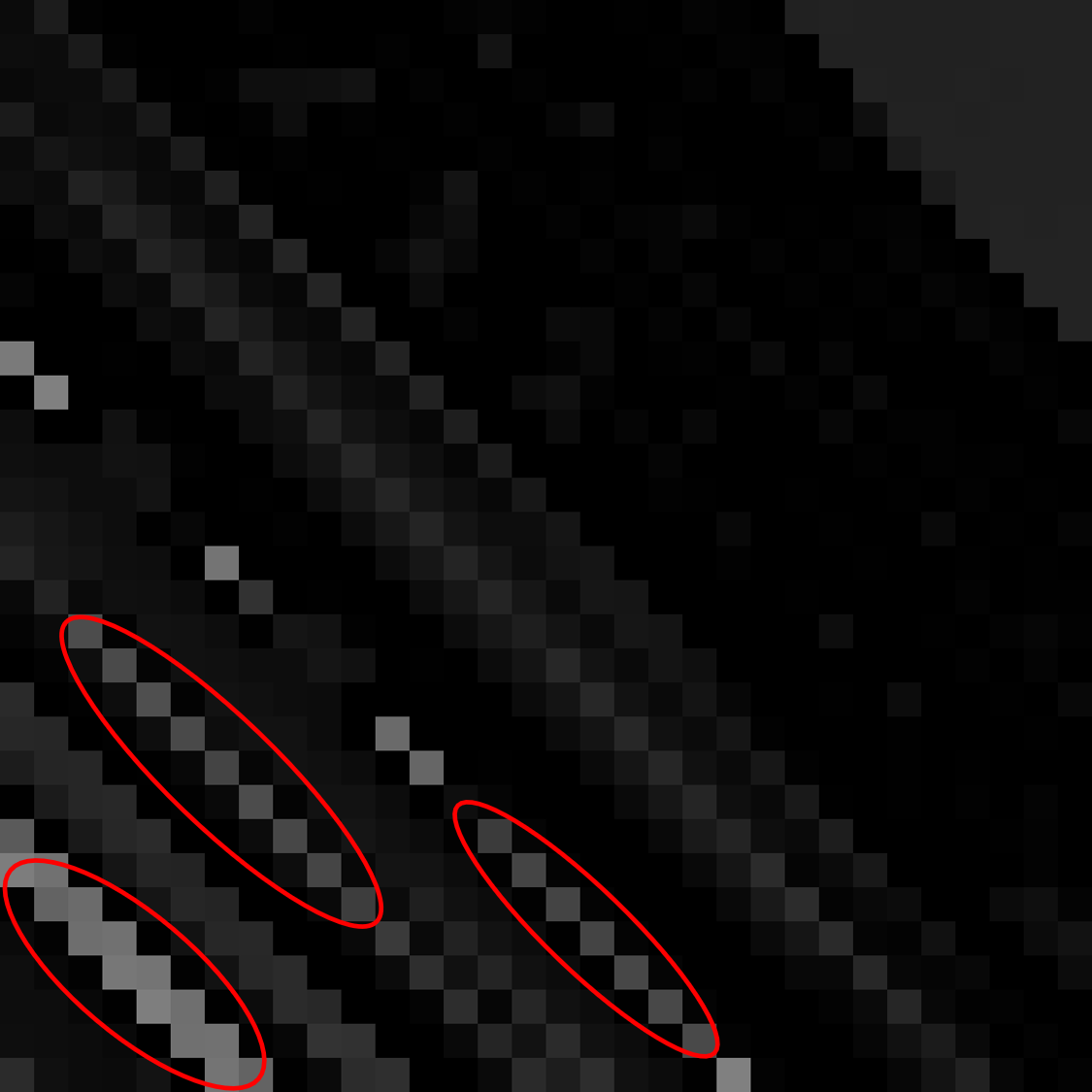}
\includegraphics[width=1.385in]{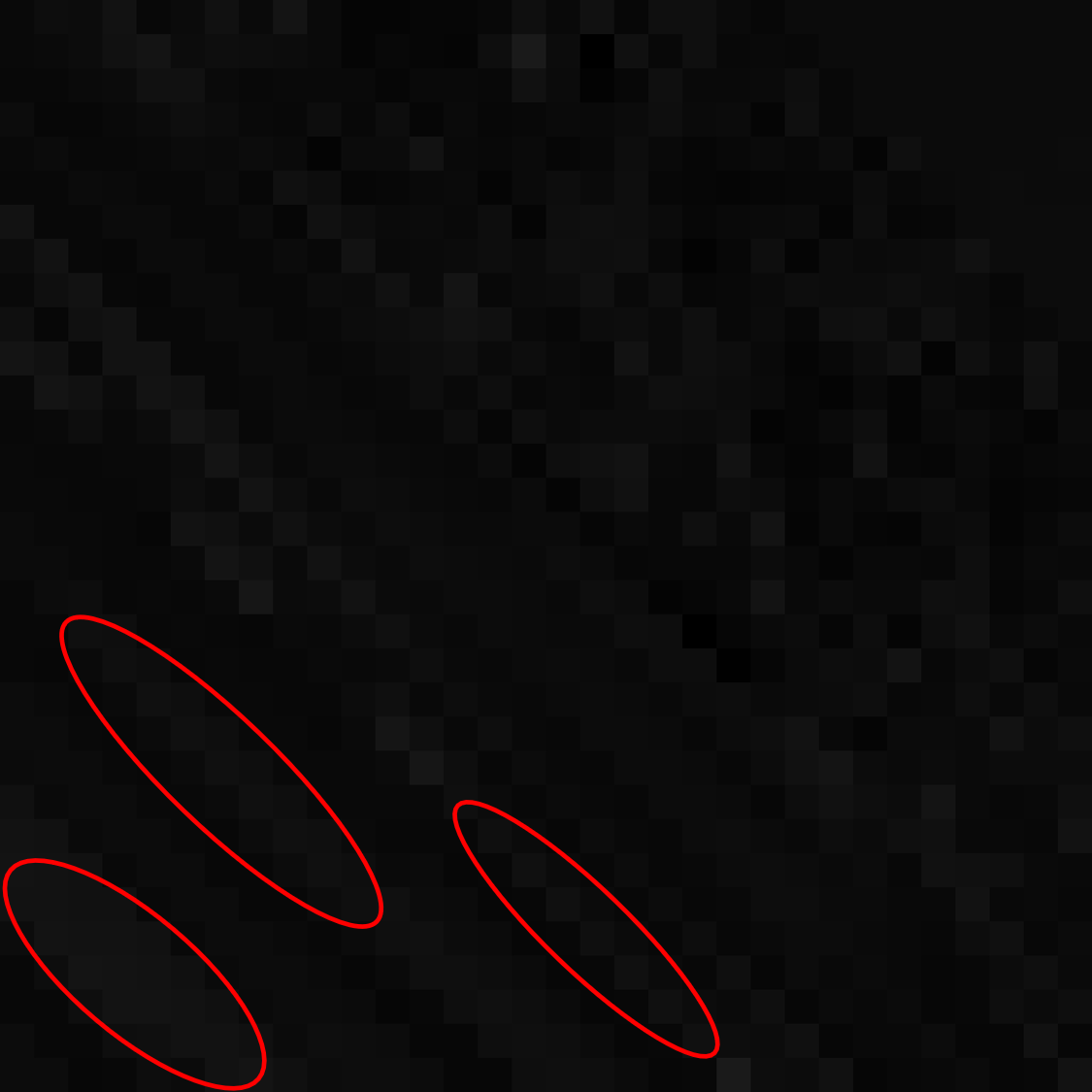}
\includegraphics[width=1.385in]{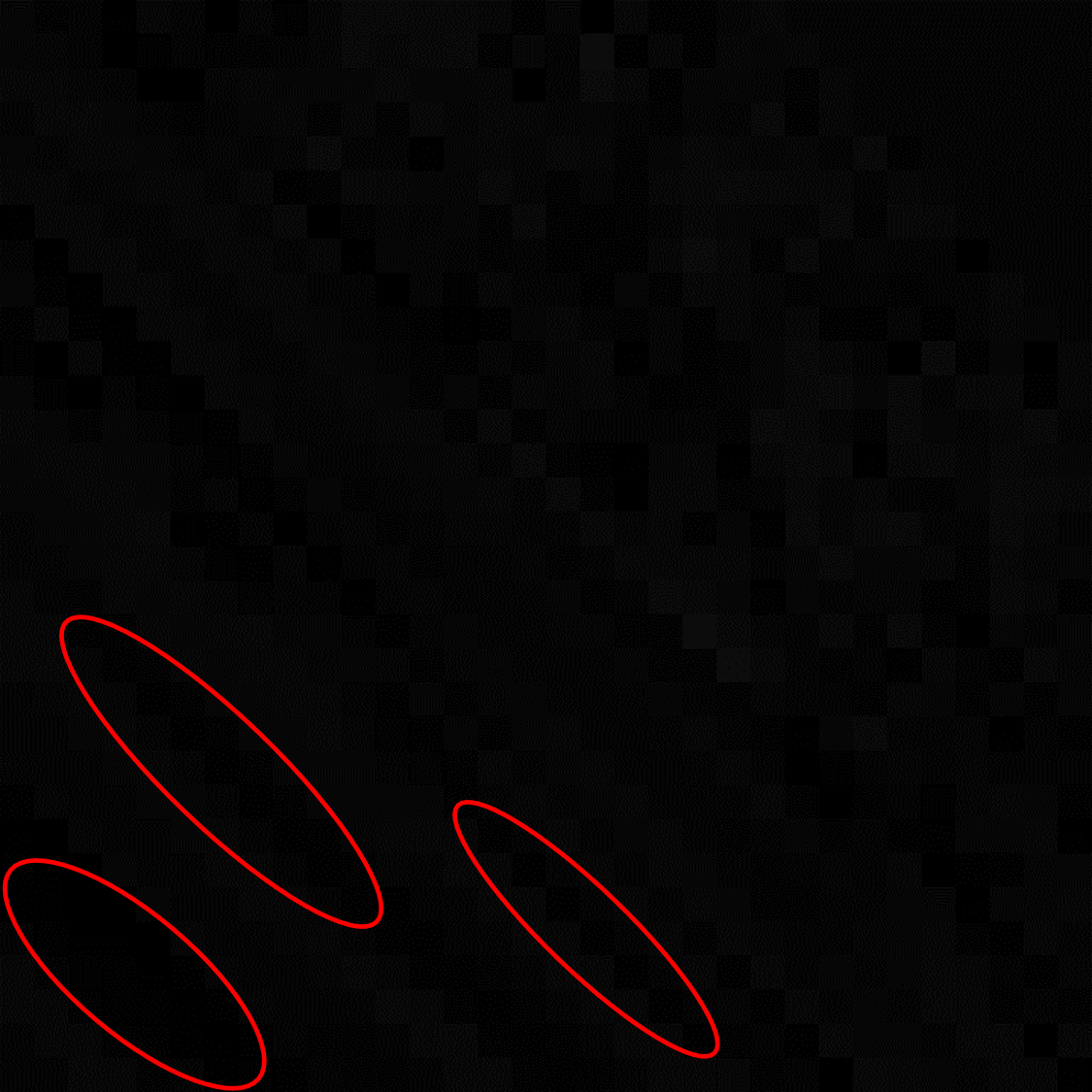}
\\
\vspace{0.02in}
\includegraphics[width=1.385in]{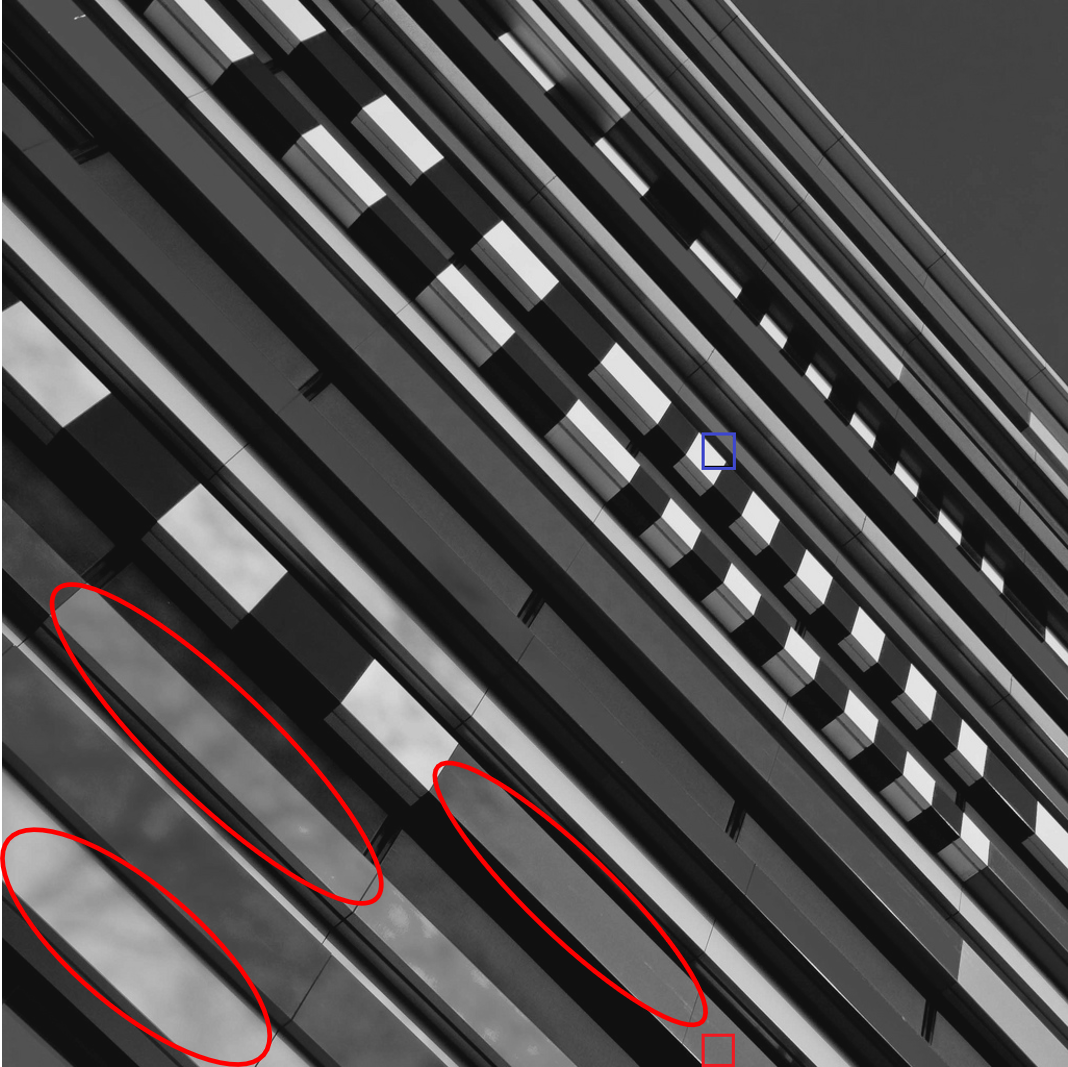}
\includegraphics[width=1.385in]{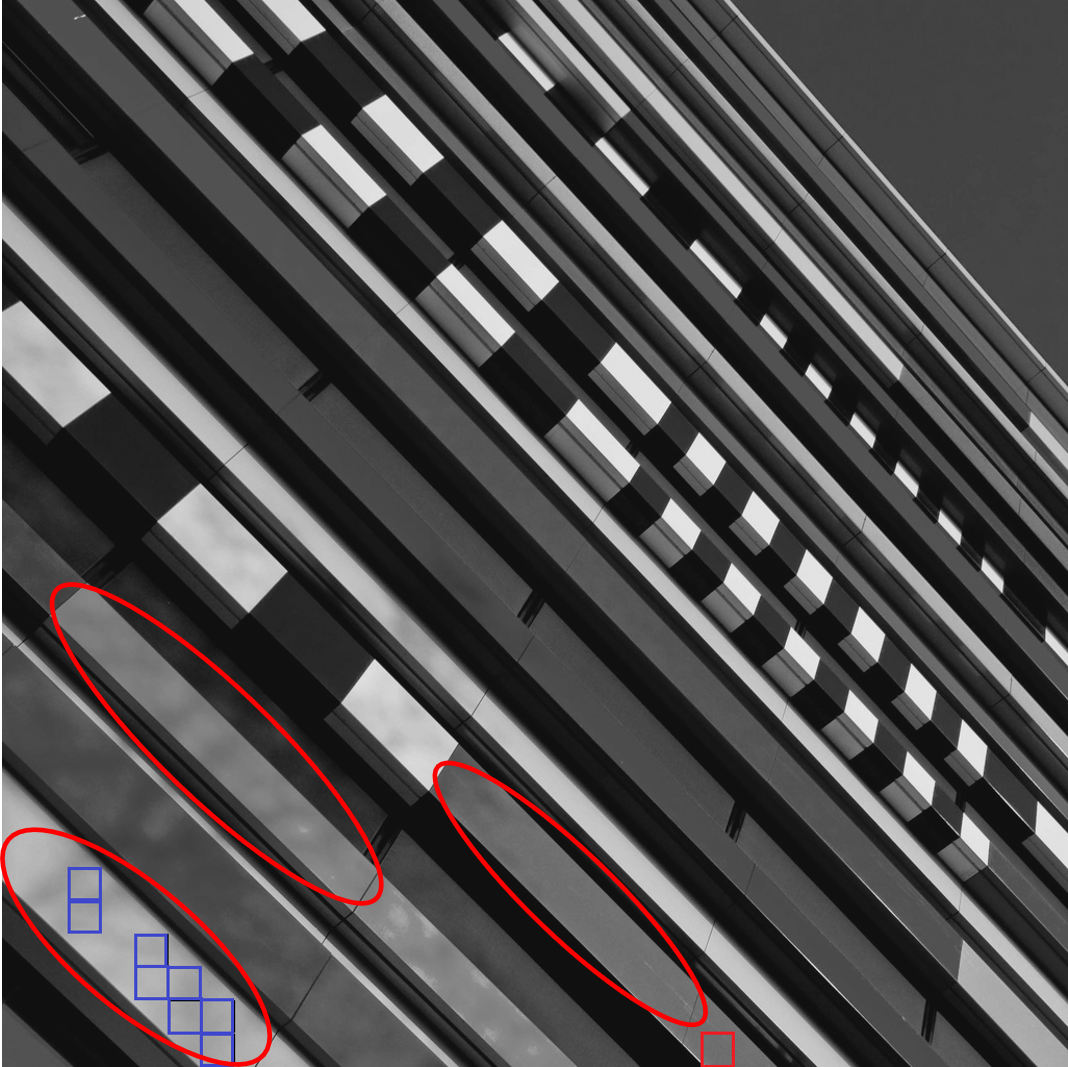}
\includegraphics[width=1.385in]{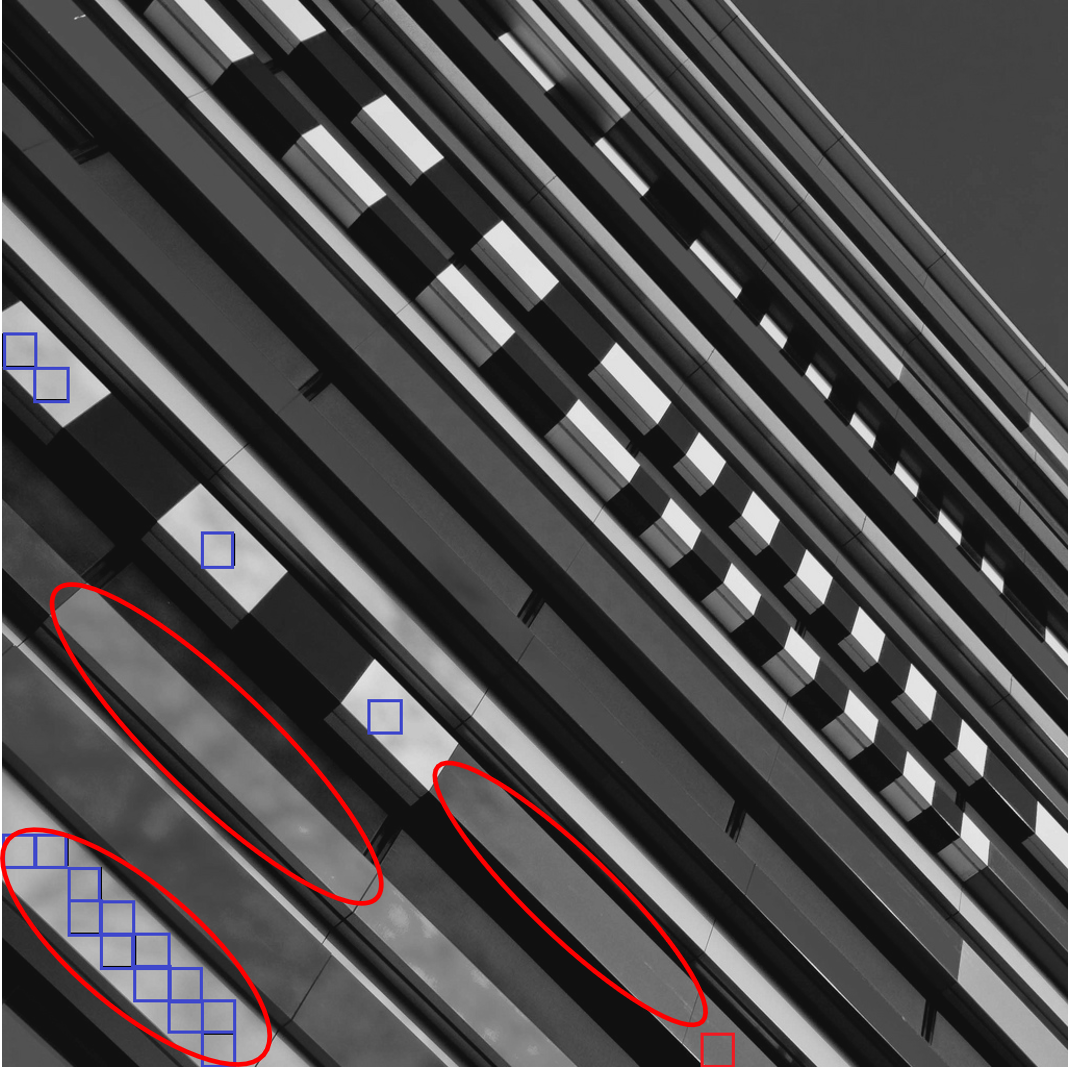}
\includegraphics[width=1.385in]{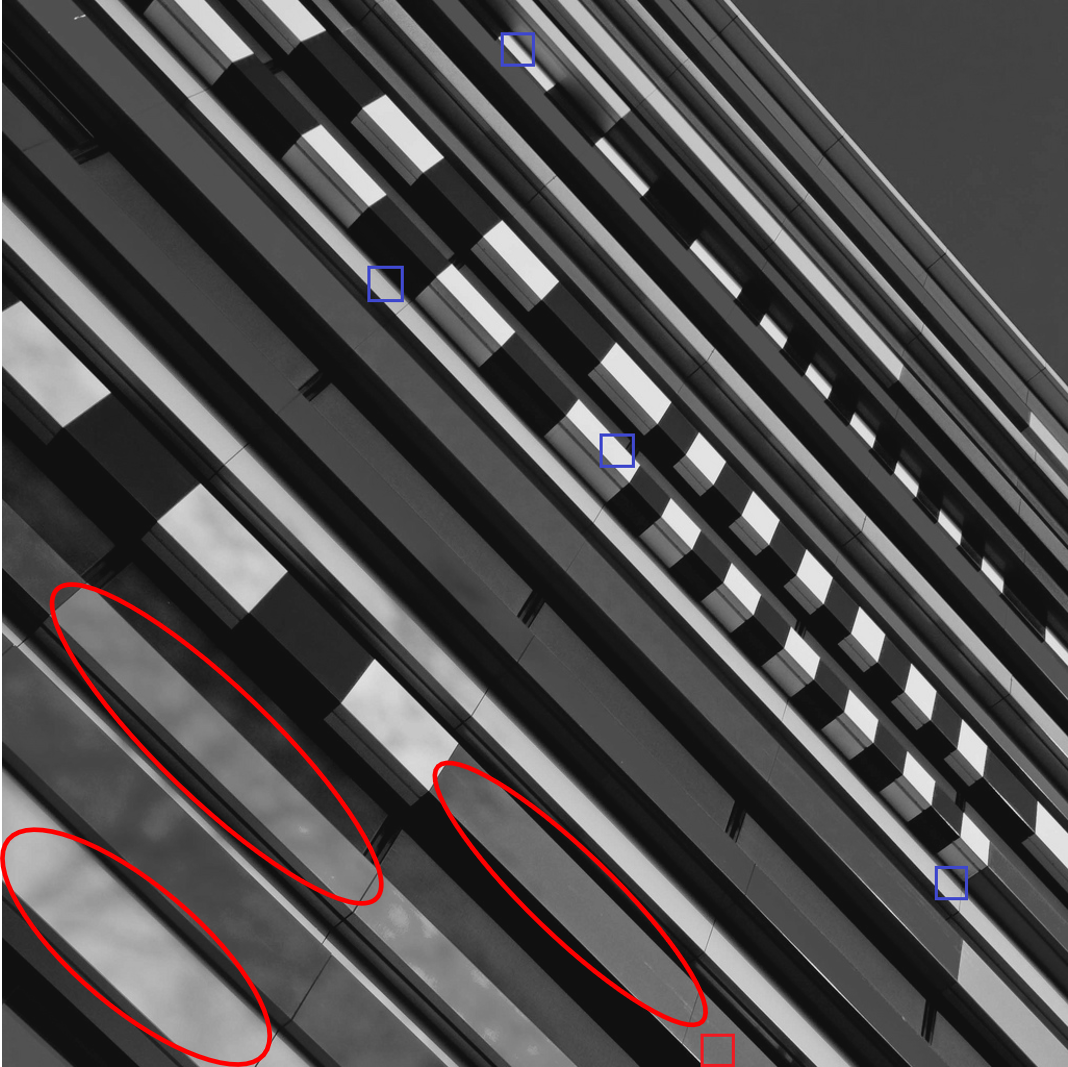}
\includegraphics[width=1.385in]{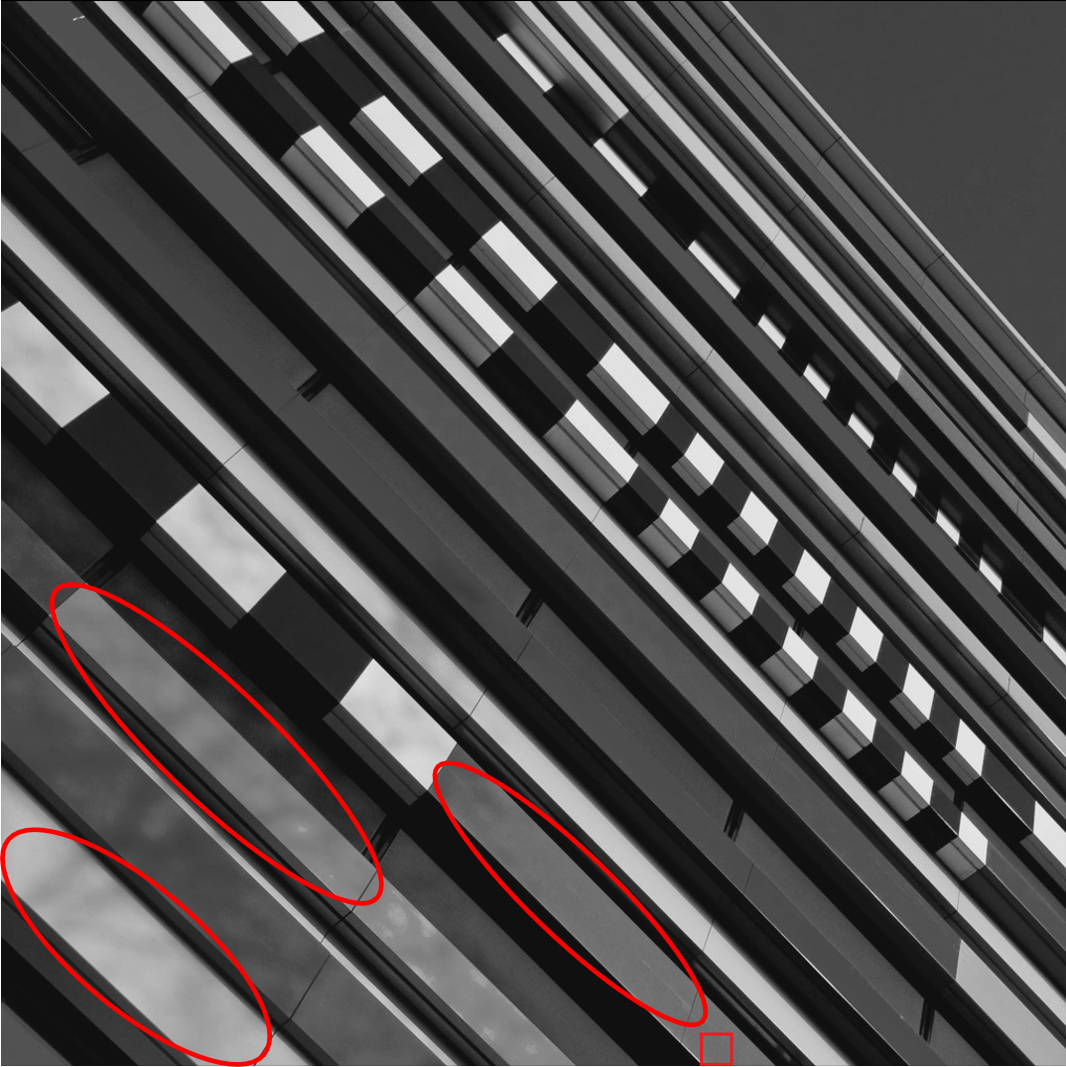}
\vskip -0.026in \tiny{$\gamma$=0 \quad\quad\quad\quad\quad\quad\quad\quad\quad\quad\quad\quad\quad\quad\quad\quad\quad\quad $\gamma$=0.001 \quad\quad\quad\quad\quad\quad\quad\quad\quad\quad\quad\quad\quad\quad\quad\quad\quad\quad $\gamma$=0.01 \quad\quad\quad\quad\quad\quad\quad\quad\quad\quad\quad\quad\quad\quad\quad\quad\quad\quad $\gamma$=0.1\quad\quad\quad\quad\quad\quad\quad\quad\quad\quad\quad\quad\quad\quad\quad\quad\quad\quad $\gamma$=1.0}

\end{center}
\vskip -0.1in
\label{fig:7}
\end{figure*}

\begin{figure*}[t]
\vskip -0.14in
\begin{center}
\includegraphics[width=1.385in]{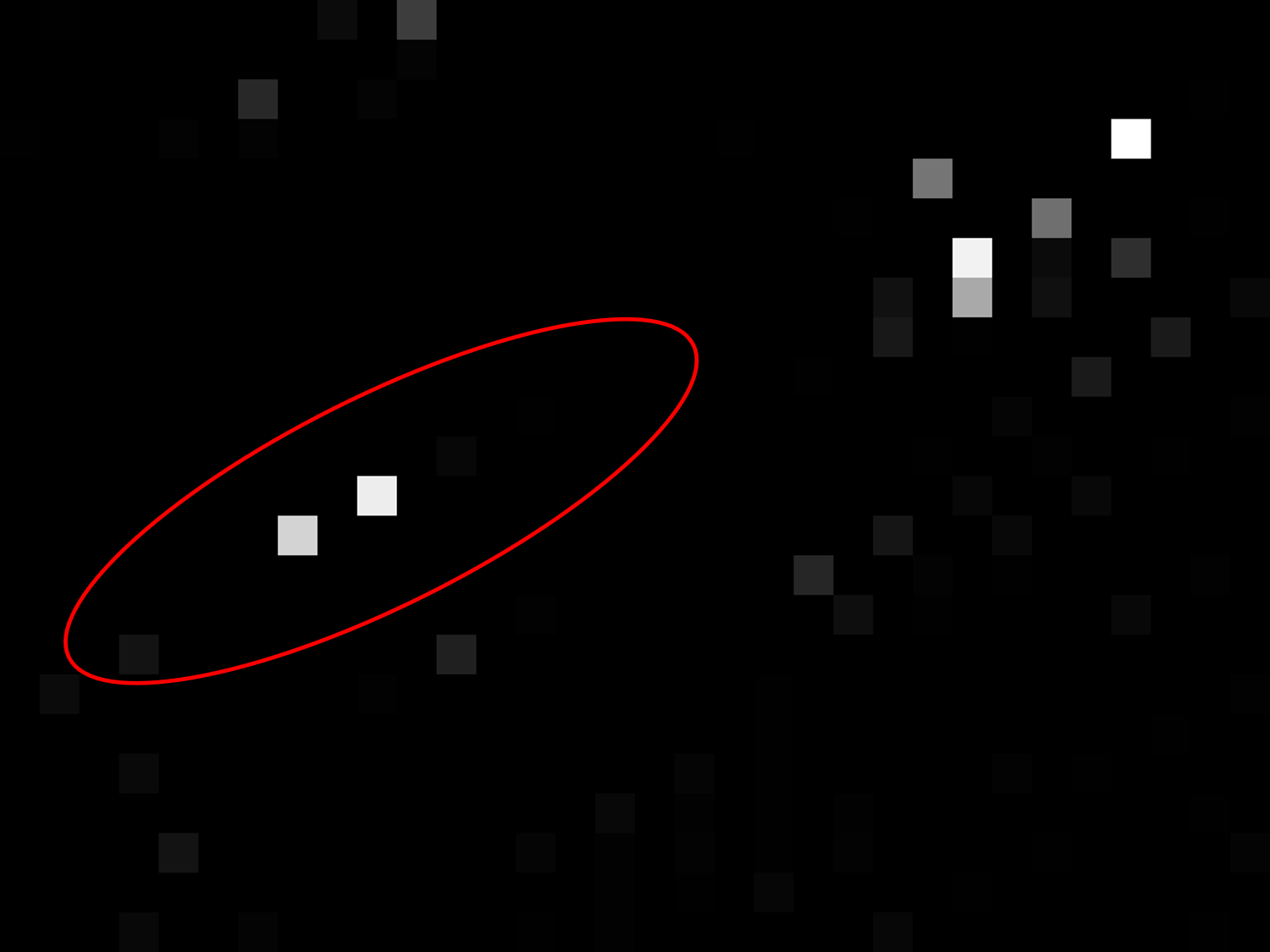}
\includegraphics[width=1.385in]{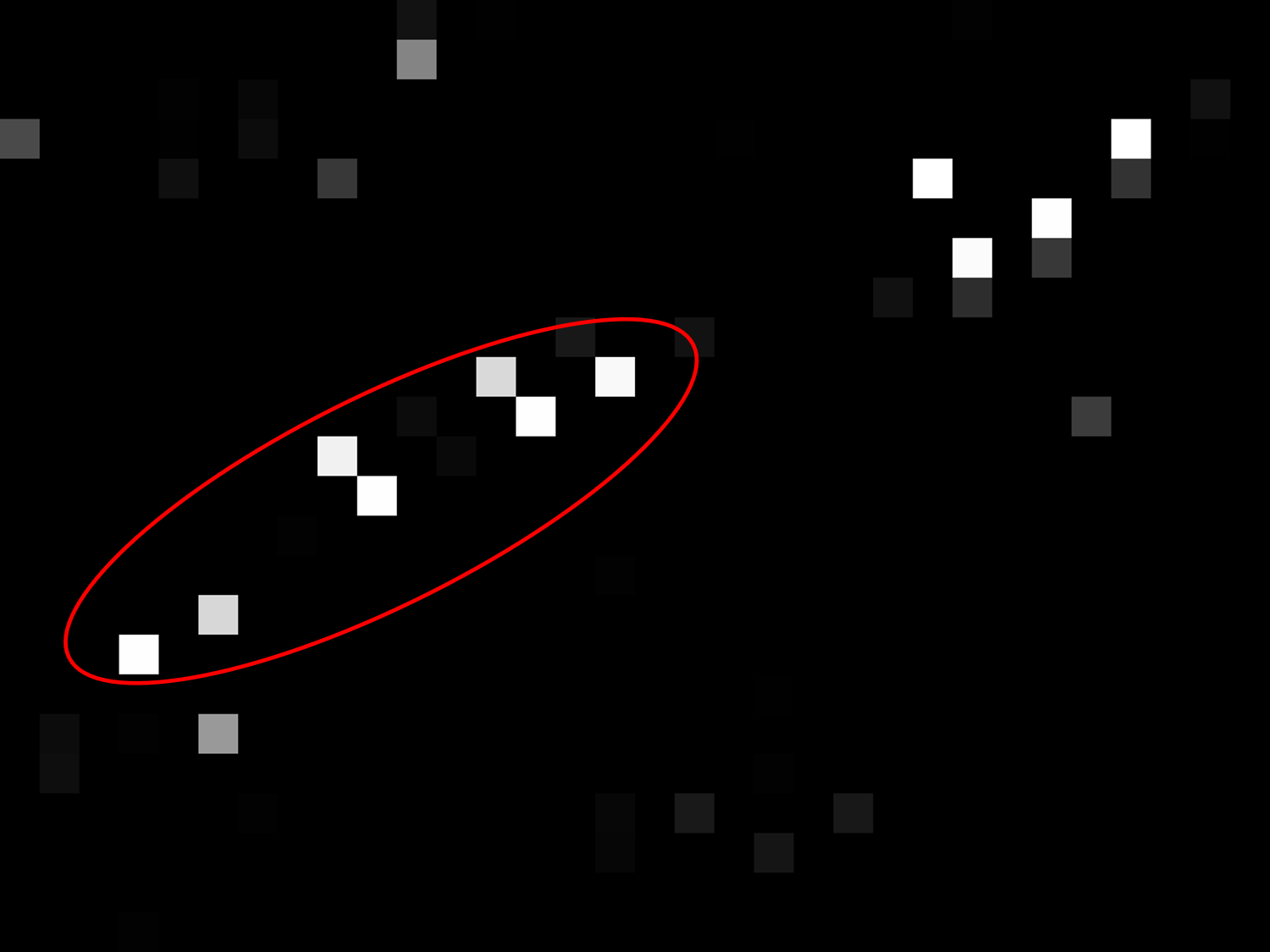}
\includegraphics[width=1.385in]{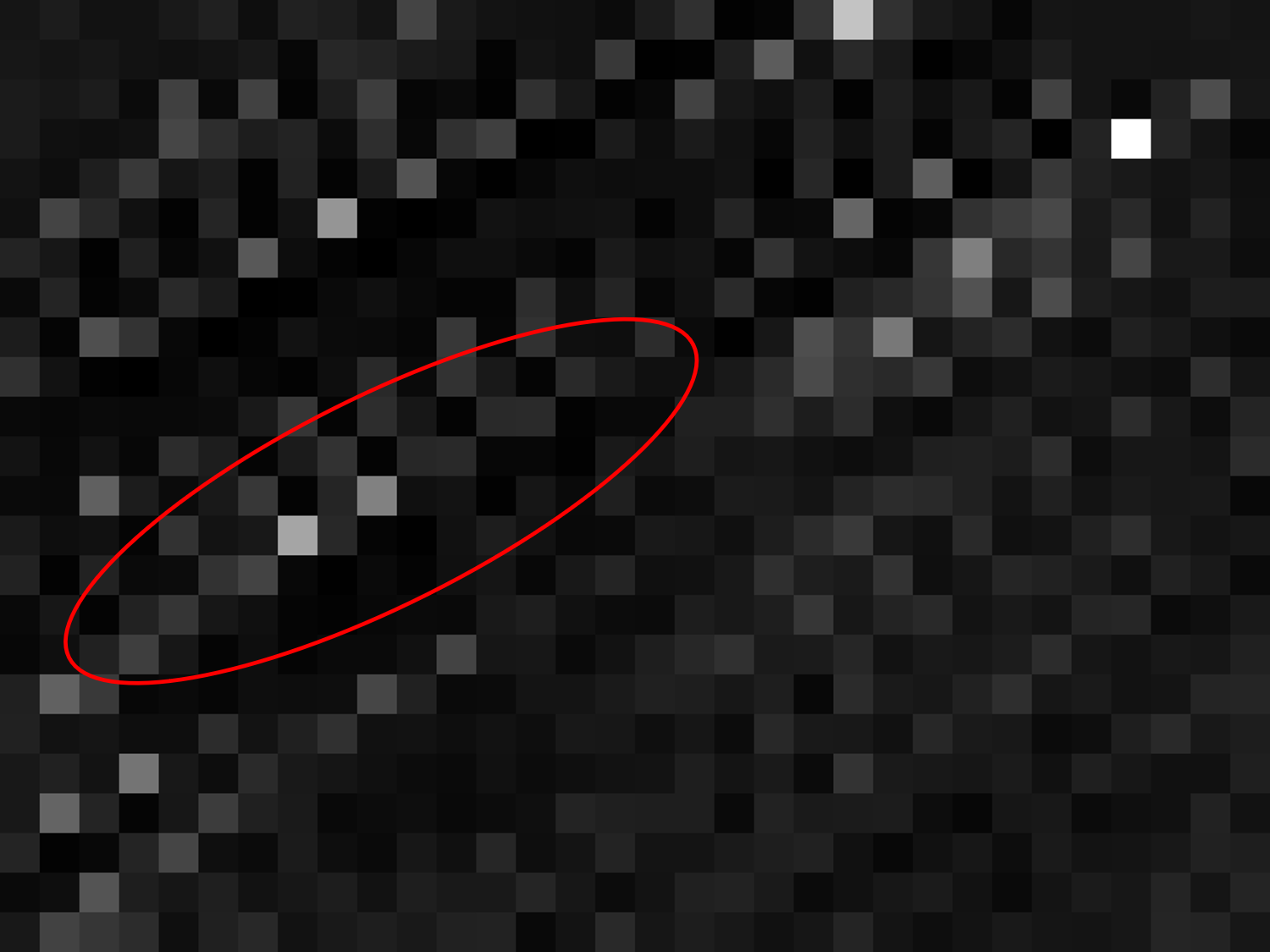}
\includegraphics[width=1.385in]{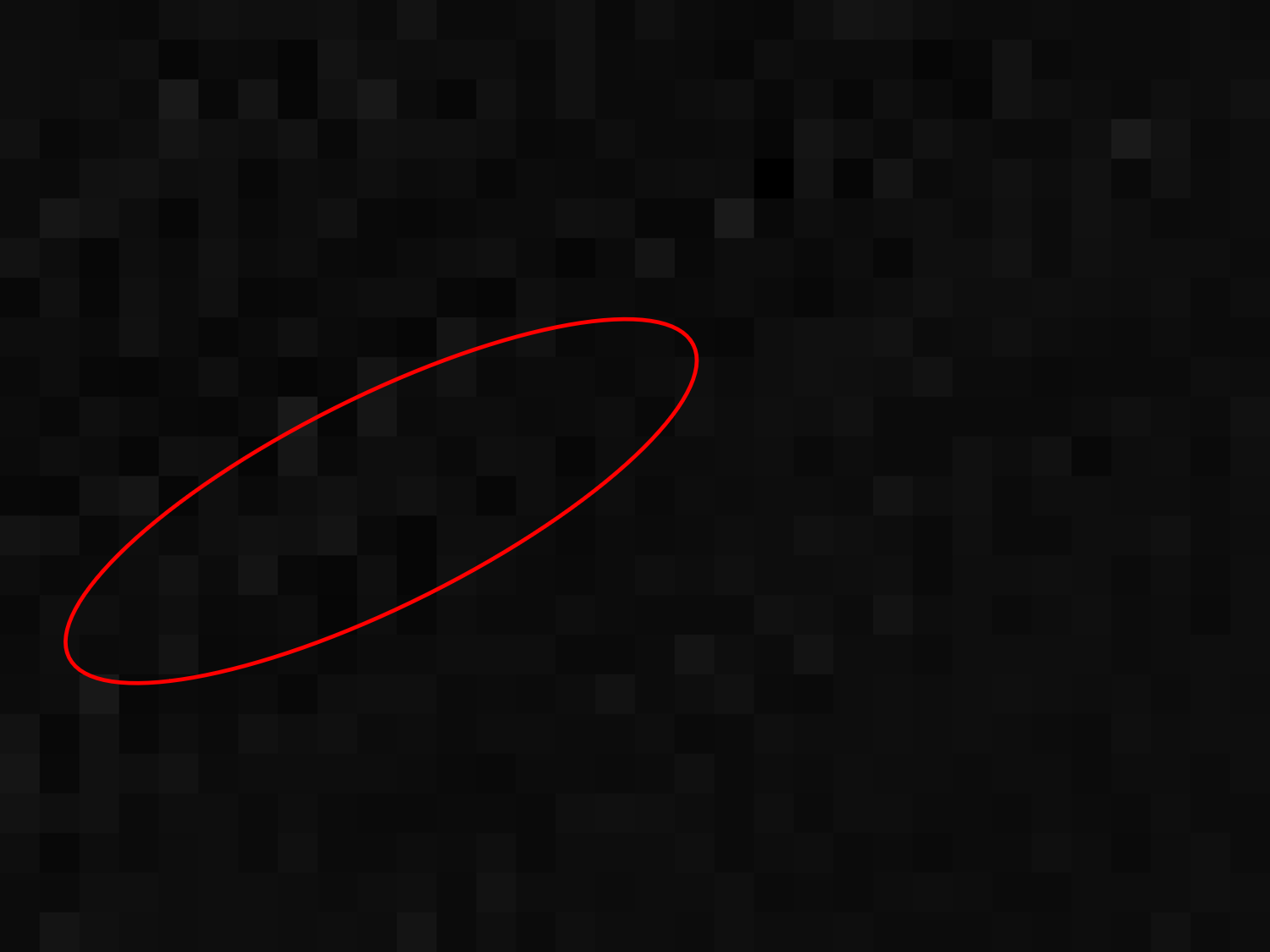}
\includegraphics[width=1.385in]{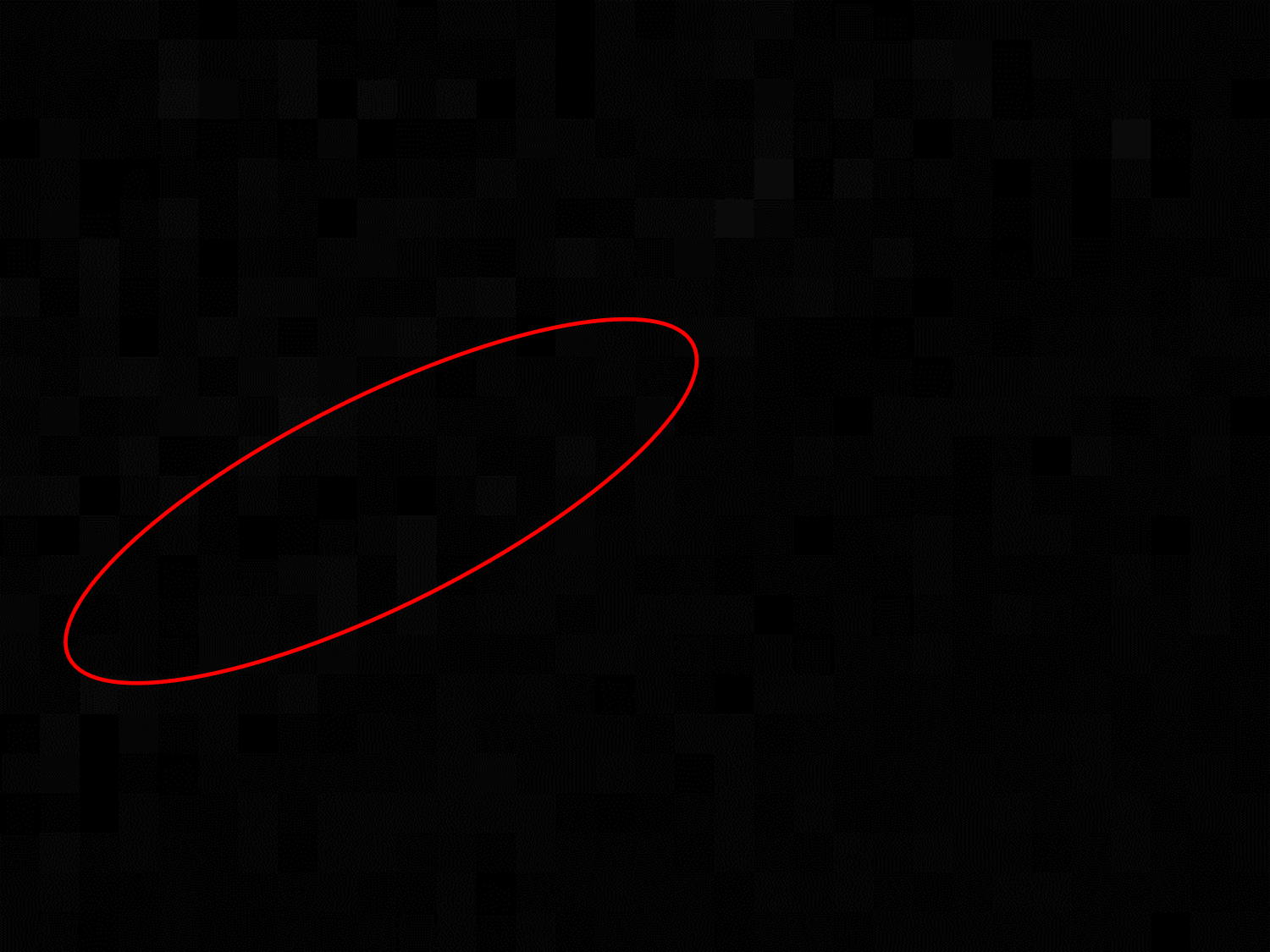}
\\
\vspace{0.02in}
\includegraphics[width=1.385in]{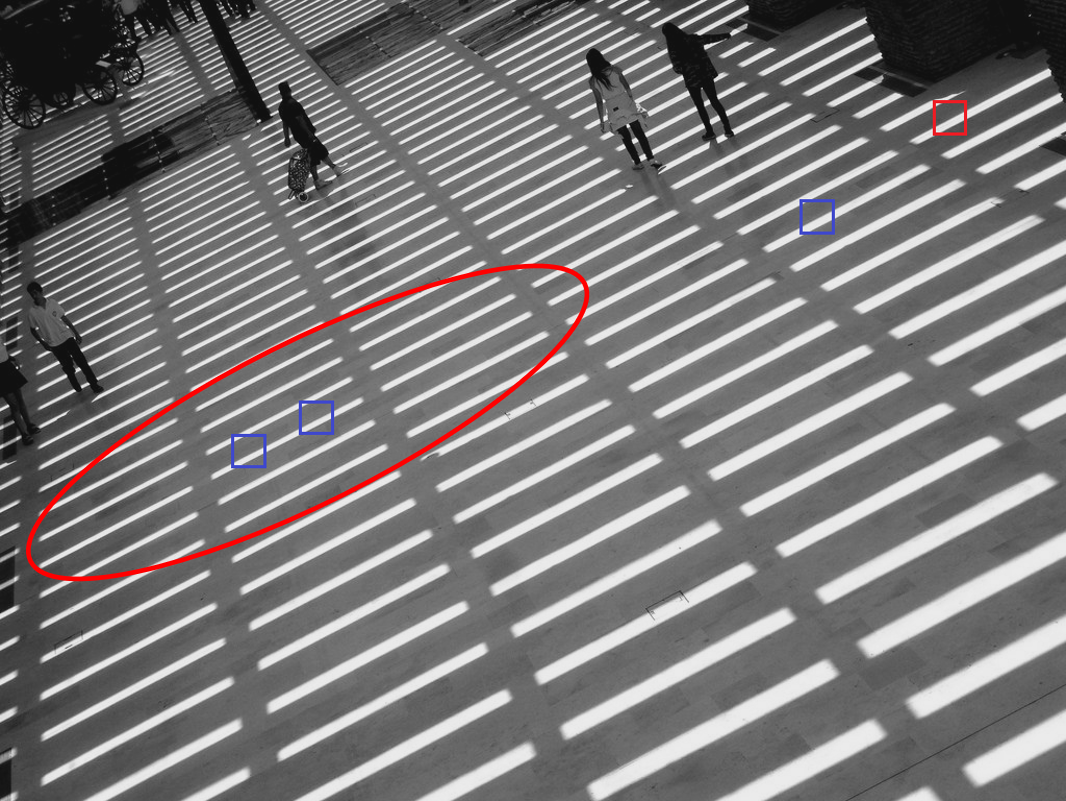}
\includegraphics[width=1.385in]{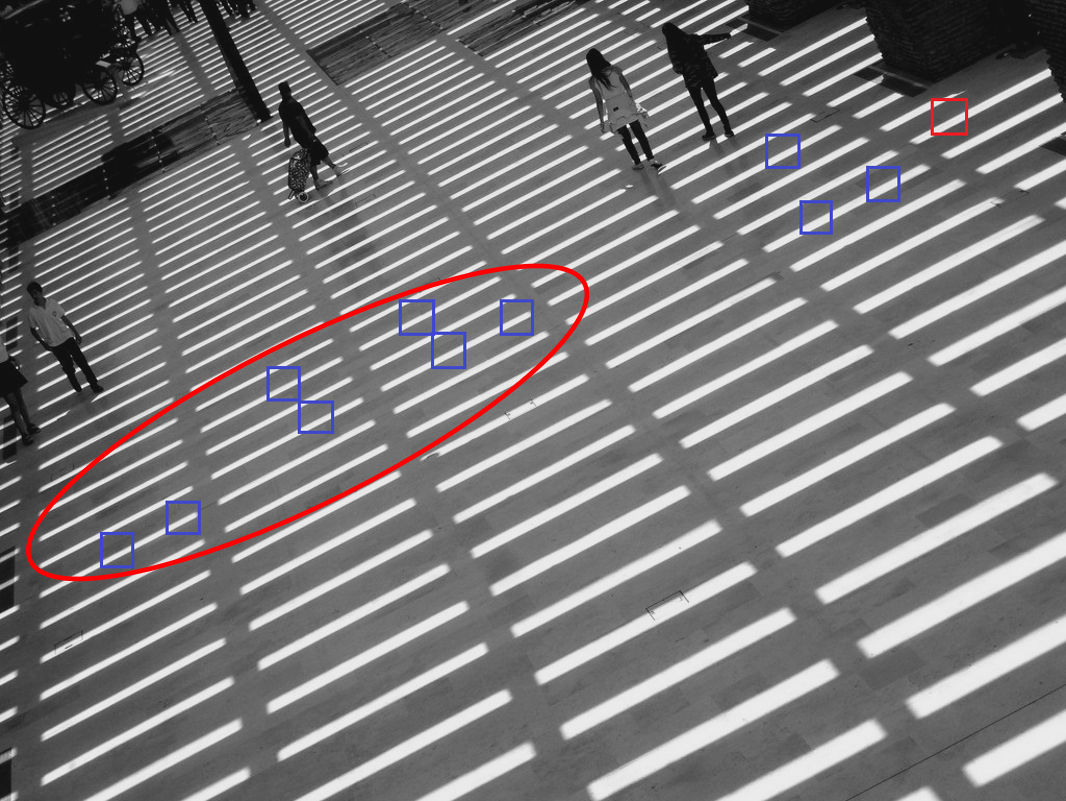}
\includegraphics[width=1.385in]{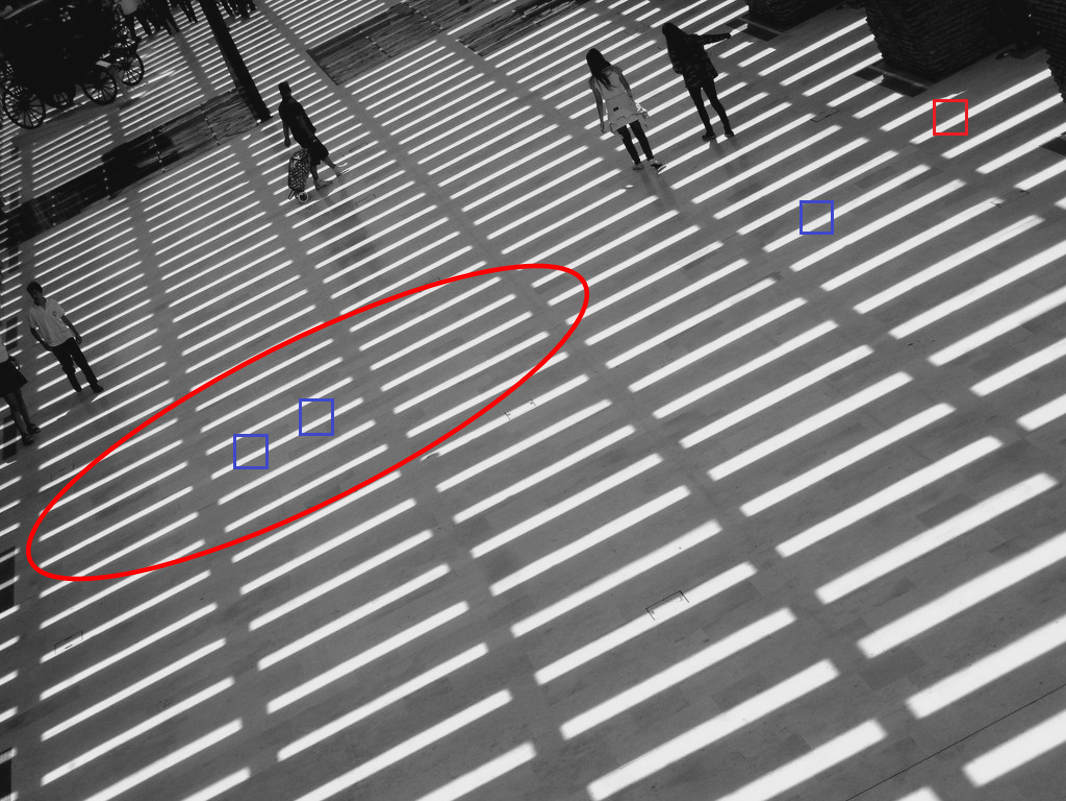}
\includegraphics[width=1.385in]{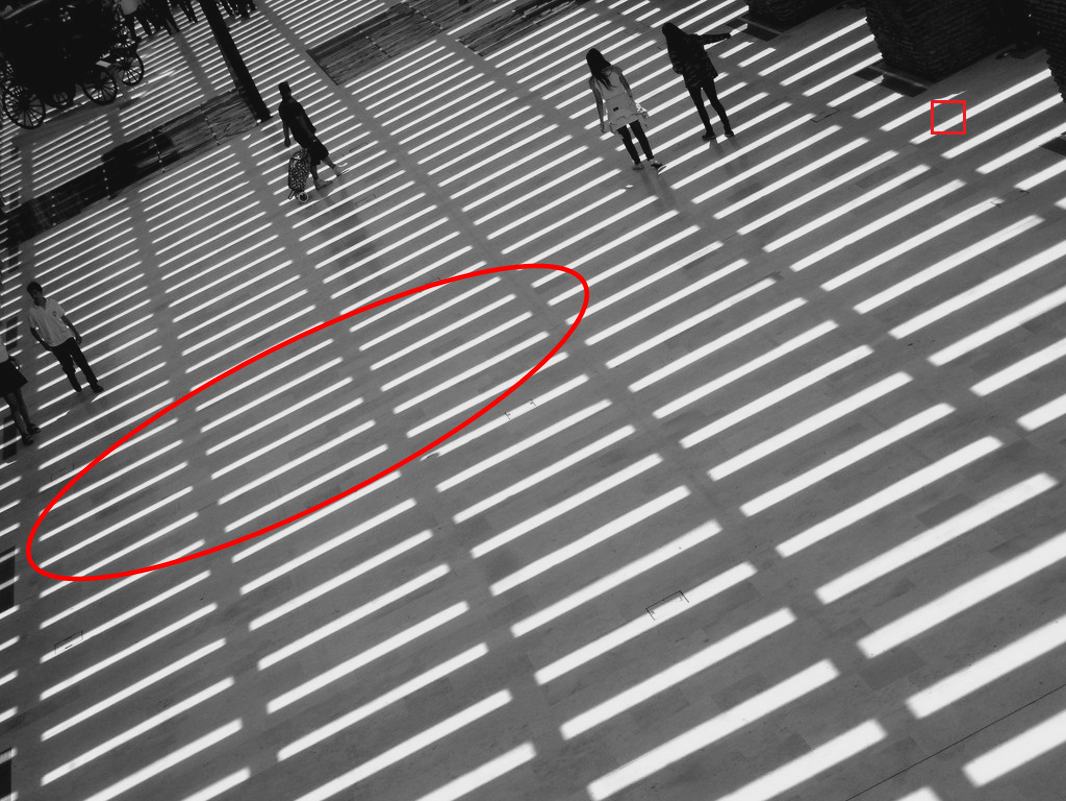}
\includegraphics[width=1.385in]{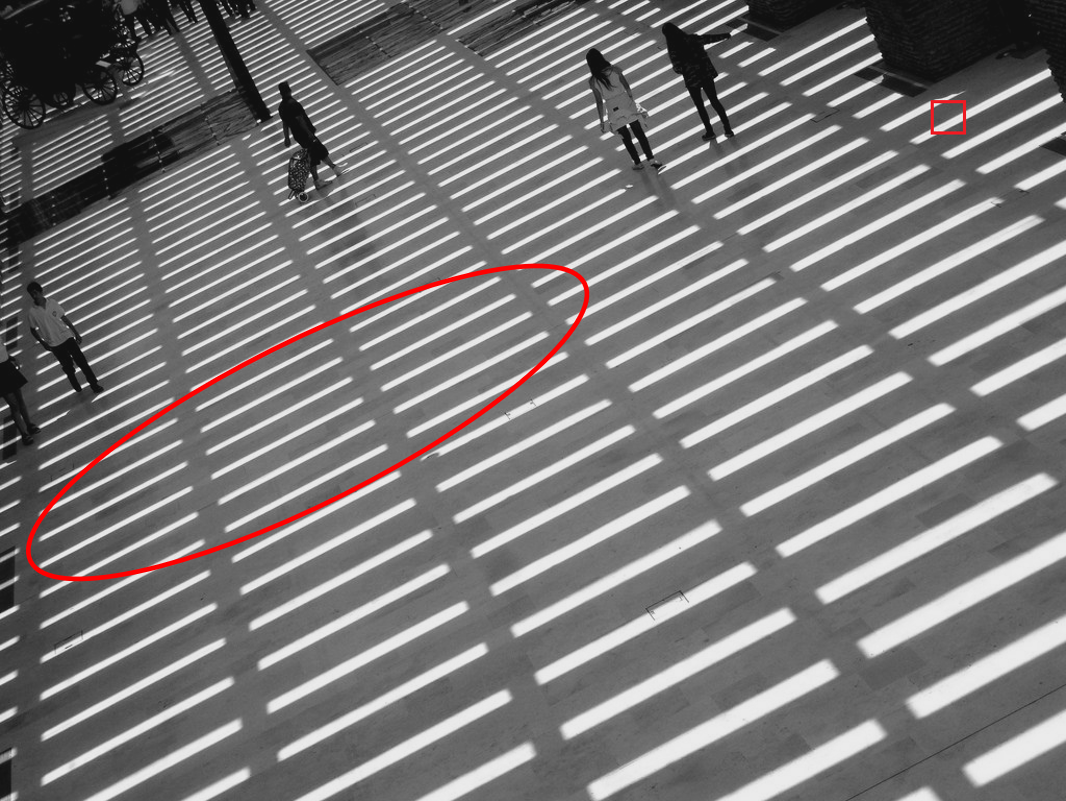}

\vskip -0.025in \tiny{$\gamma$=0 \quad\quad\quad\quad\quad\quad\quad\quad\quad\quad\quad\quad\quad\quad\quad\quad\quad\quad $\gamma$=0.001 \quad\quad\quad\quad\quad\quad\quad\quad\quad\quad\quad\quad\quad\quad\quad\quad\quad\quad $\gamma$=0.01 \quad\quad\quad\quad\quad\quad\quad\quad\quad\quad\quad\quad\quad\quad\quad\quad\quad\quad $\gamma$=0.1
\quad\quad\quad\quad\quad\quad\quad\quad\quad\quad\quad\quad\quad\quad\quad\quad\quad\quad $\gamma$=1.0}

\end{center}
\vskip -0.17in
   \caption{The visualization of the learned affinity matrix and its corresponding blocks on the original image in terms of different $\gamma$. The red blocks are the current image patches and the blue blocks are the corresponding patches mapped from the highlighted elements of affinity matrix.}
\vskip -0.2in
\label{fig:7}
\end{figure*}

\vspace{-0.08in}
\begin{equation}
\hat{y}_{(i,j)} = \frac{1}{\mathcal{C}(y)}\sum_{\forall p,q}f(y_{(i,j)}, y_{(p,q)})g(y_{(p,q)})
\label{eq:5}
\vspace{-0.07in}
\end{equation}
where $p\in \{1,2,...,w_{B}\}$ and $q\in \{1,2,...,h_{B}\}$ are the position indexes of different image blocks. The function $f$ is responsible for computing the affinities between the measurements of different image blocks. The unary function $g$ computes a representation of the input measurements $y_{(p,q)}$. From Eq.~\eqref{eq:5}, we can get that all measurements ($\forall$) of image $x$ are considered for the current measurement $y_{(i,j)}$. The factor $\mathcal{C}(y)$ is used to normalize the final response. In view of the function $f$, an extension of the gaussian function is utilized in our model to compute similarity in an embedding space. Besides, considering the coupling between the non-local information as shown in Eq.~\eqref{eq:4}, we define $f$ as
\vspace{-0.06in}
\begin{equation}
f(y_{(i,j)}, y_{(p,q)})\hspace{-0.03in}=\hspace{-0.03in}\mathop{\arg\min}\limits_{e^{\theta^{T}\phi}} \hspace{-0.03in}\|\delta_{((i,j),(p,q))}-\hspace{-0.02in}\delta_{((p,q),(i,j))}\|_{F}^{2}
\label{eq:6}
\vspace{-0.05in}
\end{equation}
where \begin{tiny}$\delta_{(\hspace{-0.01in}(i,j)\hspace{-0.01in},\hspace{-0.02in}(p,q)\hspace{-0.01in})}$=$e^{\theta(y_{(i,j)})^{T}\hspace{-0.02in}\phi(y_{(p,q)})}\hspace{-0.02in}$ \end{tiny},\hspace{-0.02in} \begin{tiny}$\delta_{(\hspace{-0.01in}(p,q)\hspace{-0.01in},\hspace{-0.02in}(i,j)\hspace{-0.01in})}$$=$$e^{\theta(y_{(p,q)})^{T}\hspace{-0.02in}\phi(y_{(i,j)})}\hspace{-0.02in}$\end{tiny}, and $\theta(y_{(i,j)})=W_{\theta}y_{(i,j)}$, $\phi(y_{(p,q)})=W_{\phi}y_{(p,q)}$ are two embeddings. $W_{\theta}$ and $W_{\phi}$ indicate two weight matrices to be learned. $T$ represents the transpose operator of a matrix. As above, we set $\mathcal{C}(y)=\sum_{\forall p,q}f(y_{(i,j)}, y_{(p,q)})$. Considering function $g$, we also use a linear embedding form: $g(y_{(p,q)})=W_{g}y_{(p,q)}$, where $W_{g}$ is a weight matrix to be learned. In detail, we use series of convolutional layers with $n$ kernels of size $1\times 1$ to optimize the learned matrix $W_{\theta}$, $W_{\phi}$ and $W_{g}$, where $n=n_{B}$ in our model. Besides, a residual structure is utilized, i.e., $\tilde{y}_{(i,j)} = y_{(i,j)}+\hat{y}_{(i,j)}$. Fig.~\ref{fig:3} shows the details of the non-local subnetwork in the measurement domain. In addition, it is worth noting that a affinity matrix $r$ is generated in the non-local subnetwork of measurement domain, which is composed of the affinities between different measurements, and the elements in the symmetrical positions represent the affinity coefficients between the current two measurements of referring to each other. After the non-local subnetwork in the measurement domain, a series of referenced ``measurements'' $\tilde{y}_{(i,j)}$ are generated.

\begin{figure*}[t]

\begin{minipage}[t]{0.135\textwidth}
\centering
\includegraphics[width=0.95in]{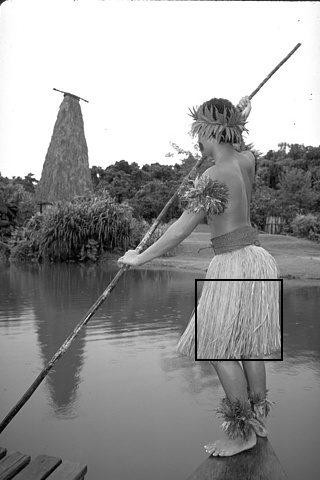}
\begin{scriptsize}

\end{scriptsize}
\end{minipage}
\hspace{-0.012in}
\begin{minipage}[t]{0.135\textwidth}
\centering
\includegraphics[width=0.95in]{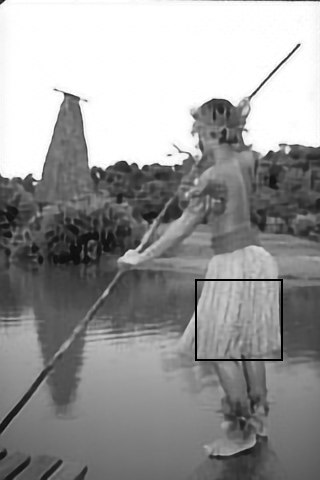}
\begin{scriptsize}

\end{scriptsize}
\end{minipage}
\hspace{-0.012in}
\begin{minipage}[t]{0.135\textwidth}
\centering
\includegraphics[width=0.95in]{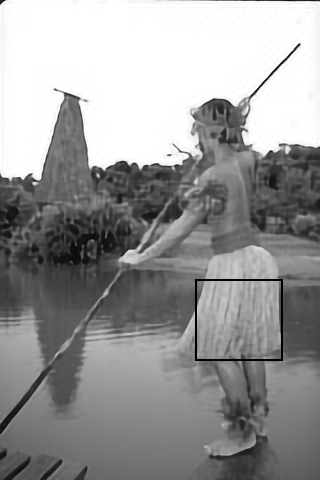}
\begin{scriptsize}

\end{scriptsize}
\end{minipage}
\hspace{-0.012in}
\begin{minipage}[t]{0.135\textwidth}
\centering
\includegraphics[width=0.95in]{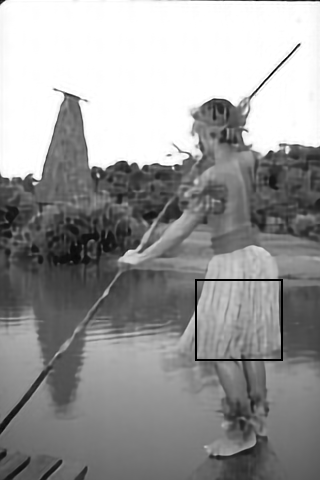}
\begin{scriptsize}

\end{scriptsize}
\end{minipage}
\hspace{-0.012in}
\begin{minipage}[t]{0.135\textwidth}
\centering
\includegraphics[width=0.95in]{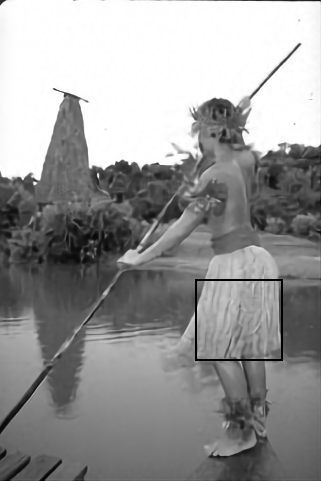}
\begin{scriptsize}

\end{scriptsize}
\end{minipage}
\hspace{-0.012in}
\begin{minipage}[t]{0.135\textwidth}
\centering
\includegraphics[width=0.95in]{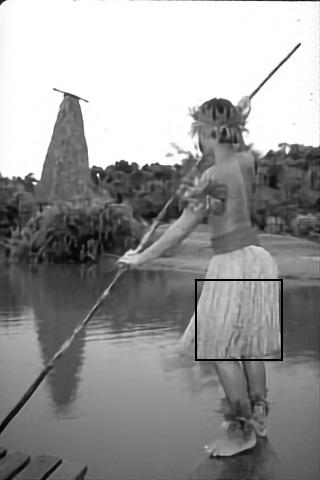}
\begin{scriptsize}

\end{scriptsize}
\end{minipage}
\hspace{-0.012in}
\begin{minipage}[t]{0.135\textwidth}
\centering
\includegraphics[width=0.95in]{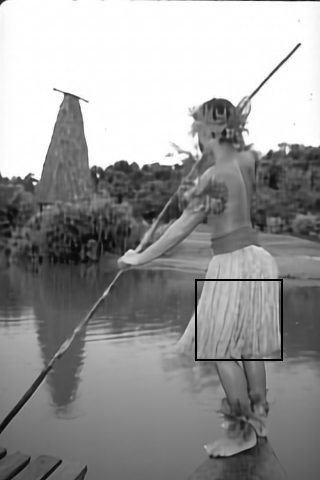}
\begin{scriptsize}

\end{scriptsize}
\end{minipage}
\vspace{-0.13in}
\\
\begin{minipage}[t]{0.135\textwidth}
\centering
\includegraphics[width=0.95in]{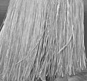}
\begin{scriptsize}
\centering
\vskip -0.52 cm \begin{tiny}Methods$\backslash$PSNR$\backslash$SSIM\end{tiny}
\end{scriptsize}
\end{minipage}
\hspace{-0.012in}
\begin{minipage}[t]{0.135\textwidth}
\centering
\includegraphics[width=0.95in]{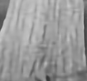}
\begin{scriptsize}
\centering
\vskip -0.52 cm \begin{tiny}CSNet~\cite{ref55}$\backslash$27.70$\backslash$0.8245\end{tiny}
\end{scriptsize}
\end{minipage}
\hspace{-0.012in}
\begin{minipage}[t]{0.135\textwidth}
\centering
\includegraphics[width=0.95in]{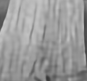}
\begin{scriptsize}
\centering
\vskip -0.38 cm \begin{tiny}CSNet$^{+}$~\cite{ref6}$\backslash$27.63$\backslash$0.8266\end{tiny}

\end{scriptsize}
\end{minipage}
\hspace{-0.012in}
\begin{minipage}[t]{0.135\textwidth}
\centering
\includegraphics[width=0.95in]{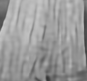}
\begin{scriptsize}
\centering
\vskip -0.52 cm \begin{tiny}SCSNet~\cite{ref19}$\backslash$27.72$\backslash$0.8307\end{tiny}
\end{scriptsize}
\end{minipage}
\hspace{-0.012in}
\begin{minipage}[t]{0.135\textwidth}
\centering
\includegraphics[width=0.95in]{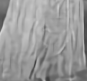}
\begin{scriptsize}
\centering
\vskip -0.52 cm \begin{tiny}OPINENet$^{+}$~\cite{ref23}$\backslash$28.68$\backslash$0.8571\end{tiny}
\end{scriptsize}
\end{minipage}
\hspace{-0.012in}
\begin{minipage}[t]{0.135\textwidth}
\centering
\includegraphics[width=0.95in]{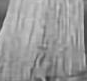}
\begin{scriptsize}
\centering
\vskip -0.52 cm \begin{tiny}AMP-Net~\cite{ref22}$\backslash$28.71$\backslash$0.8466\end{tiny}
\end{scriptsize}
\end{minipage}
\hspace{-0.012in}
\begin{minipage}[t]{0.135\textwidth}
\centering
\includegraphics[width=0.95in]{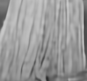}
\begin{scriptsize}
\centering
\vskip -0.52 cm \begin{tiny}NL-CSNet$^{\ast}$$\backslash$\textbf{28.76}$\backslash$\textbf{0.8672}\end{tiny}
\end{scriptsize}
\end{minipage}

\vspace{-0.13in} \caption{Visual quality comparisons of deep network-based CS methods using learned sampling matrix on one sample image from BSD68 in case of sampling rate = 0.1.}

\label{fig:8}
\vspace{-0.18in}
\end{figure*}

In order to reconstruct the initial reconstruction ($\hat{x}$), an upsampling operation is subsequently performed by $\Phi_{B}\tilde{y}_{(i,j)}$, where $\Phi_{B}$ is a $B^{2}\times n_{B}$ matrix. In our model, we utilize a convolutional layer with $B^{2}$ filters of size 1$\times$1$\times n_{B}$ to learn the upsampling matrix $\Phi_{B}$ (shown in Fig.~\ref{fig:2}(b)), after which a series of vectors of size $1\times 1\times B^{2}$ are generated. Finaly, two additional operators, i.e., reshape and concatenation ($R_{t}$)~\cite{ref6}, are appended as shown in Fig.~\ref{fig:2}(c) to obtain the initial reconstruction. By exploring correlations between the measurements of different image blocks, a better initial reconstruction is usually obtained, which is more favored for the subsequent deep reconstruction. Through the above analysis, we can obtain that the non-local model in the measurement domain builds a coarse-grained reference between the measurements of non-overlapping image blocks. Fig.~\ref{fig:1} shows the visual results of the learned affinity matrices in the measurement domain, from which we can observe that our non-local subnetwork is capable of exploring non-local patches with similar structures in the whole image space.

\vspace{-0.1in}
\subsection{Non-Local Subnetwork in Multi-Scale Feature Domain}
\vspace{-0.01in}

After the initial reconstruction, the non-local subnetwork in multi-scale feature domain (dubbed MS-NLNet) is appended to further enhance the CS reconstruction. For the structure of MS-NLNet, a series of horizontal and vertical branches are included as shown in Fig.~\ref{fig:4} to form a grid architecture~\cite{ref68}. Specifically, the horizontal branch is responsible for the feature extraction and the non-local knowledge exploiting under a certain scale space. The vertical branch mainly focuses on the transformation of different scale spaces, which includes the downsampling and upsampling branches that separately responsible for the down and up sampling operations of the intermediate feature maps. For simplicity, the number of the downsampling and upsampling branches is set as $d_{u}$ and $d_{v}$ respectively in our model.

In view of the internal structure details of MS-NLNet, three types of submodules, i.e., downsample, upsample and non-local submodule, are developed. For the first two submodules, the dense connection of the proposed residual blocks (shown in Fig.~\ref{fig:4}) is utilized and the number of the residual blocks is set as $d_{l}$ and $d_{r}$ respectively in these two types of submodules. In addition, it is worth noting that an additional aggregation operator (Conv $1\times 1$) is appended to aggregate the input feature maps together in these two types of submodules. Their structure details are shown in the subgraphs (a) and (b) of Fig.~\ref{fig:4}. For the non-local submodule, we establish a non-local reference between the feature points (with full channels) in the multi-scale space. For simplicity, we set the scale number of our MS-NLNet as $S_{B}$ and there are $S_{N}$ non-local submodules in each scale space. Therefore, the non-local operation in the multi-scale feature domain can be expressed as:

\vspace{-0.1in}
\begin{equation}
\hat{z}_{(i,j)}^{s_{t}} = \frac{1}{\mathcal{C}(z^{s_{t}})}\sum_{\forall p,q}f(z_{(i,j)}^{s_{t}}, z_{(p,q)}^{s_{t}})g(z_{(p,q)}^{s_{t}})
\vspace{-0.06in}
\end{equation}
where $\vspace{-0.02in}z_{(i,j)}^{s_{t}}$ indicates the current feature points (with full channels) of the input feature maps in the $t$-$th$ non-local submodule of the $s$-$th$ scale space, where $t=\{1,2,...S_{N}\}$ and $s=\{1,2,...S_{B}\}$. $(i, j)$ and $(p, q)$ are the position indexes of the feature points. It is clear that for the current feature point $z_{(i,j)}^{s_{t}}$, all feature points ($\forall$) in the current feature maps are concerned to generate the corresponding referenced feature point $\hat{z}_{(i,j)}^{s_{t}}$. In addition, a residual structure is utilized, i.e., $\tilde{z}_{(i,j)}^{s_{t}} = z_{(i,j)}^{s_{t}}+\hat{z}_{(i,j)}^{s_{t}}$. The structure details of the non-local subnetwork in the multi-scale feature domain are shown in Fig.~\ref{fig:4}. For the functions $f$ and $g$, they have the similar definitions as that in the measurement domain (shown in Eq.~\eqref{eq:6}), i.e.,
\vspace{-0.06in}
\begin{equation}
f(z_{(i,j)}^{s_{t}}, z_{(p,q)}^{s_{t}})\hspace{-0.03in}=\hspace{-0.03in}\mathop{\arg\min}\limits_{e^{\theta^{T}\phi}} \hspace{-0.03in}\|\delta_{((i,j),(p,q))}^{s_{t}}-\hspace{-0.02in}\delta_{((p,q),(i,j))}^{s_{t}}\|_{F}^{2}
\label{eq:66}
\vspace{-0.05in}
\end{equation}
where \begin{tiny}$\delta_{(\hspace{-0.01in}(i,j)\hspace{-0.01in},\hspace{-0.02in}(p,q)\hspace{-0.01in})}^{s_{t}}$=$e^{\theta(z_{(i,j)}^{s_{t}})^{T}\hspace{-0.02in}\phi(z_{(p,q)}^{s_{t}})}\hspace{-0.02in}$ \end{tiny},\hspace{-0.02in} \begin{tiny}$\delta_{(\hspace{-0.01in}(p,q)\hspace{-0.01in},\hspace{-0.02in}(i,j)\hspace{-0.01in})}^{s_{t}}$$=$$e^{\theta(z_{(p,q)}^{s_{t}})^{T}\hspace{-0.02in}\phi(z_{(i,j)}^{s_{t}})}\hspace{-0.02in}$\end{tiny}, and $\theta(y_{(i,j)})=W_{\theta}y_{(i,j)}$, $\phi(y_{(p,q)})=W_{\phi}y_{(p,q)}$, $g(y_{(p,q)})=W_{g}y_{(p,q)}$. $W_{\theta}$, $W_{\phi}$ and $W_{g}$ indicate the weight matrices to be learned. Similar to the non-local subnetwork in the measurement domain, we use series of convolutional layers with $n$
kernels of size 1 $\times$ 1 to optimize these learned matrices, where $n$ is equal to the channel number of the input feature maps. Analogically, $\mathcal{C}(z^{s_{t}})=$$\sum_{\forall p,q}f(z_{(i,j)}^{s_{t}}, z_{(p,q)}^{s_{t}})$ is used to normalize the final response.

In fact, with the dimensional increase of the input signal, the resource consumptions of the non-local neural networks increase rapidly~\cite{ref65}. In order to alleviate this problem, a downsampling operator ($\downarrow$ in Fig.~\ref{fig:4}) is used in our non-local neural network. Besides, it is worth noting that an affinity matrix is also generated in each non-local submodule, i.e., $r^{s_{t}}$ ,which is composed of the affinities between different feature points, and the elements in the symmetrical positions represent the affinity coefficients between two feature points of referring to each other. Fig.~\ref{fig:6} shows the visual results of the learned affinity matrices in the multi-scale feature domain, from which we can observe that the non-local textures with the similar structures are explored efficiently. Compared to the non-local subnetwork in the measurement domain, the non-local subnetwork in the multi-scale feature domain builds a fine-grained reference between the dense feature representations of the feature maps.

\begin{table}[b]
\vspace{-0.16in}
  \caption{The experimental results (PSNR) for analyzing the contributions of different functional modules in terms of various sampling rates on dataset BSD68.}
  \label{tab:7}
  \vspace{-0.1in}

  \begin{tabular}{p{0.84cm}<{\centering}  p{0.57cm}<{\centering}  p{0.57cm}<{\centering}  p{0.6cm}<{\centering} |  p{0.70cm}<{\centering}  p{0.70cm}<{\centering}  p{0.70cm}<{\centering}  p{0.70cm}<{\centering}}
    \toprule

    \footnotesize{Coupling}&\footnotesize{NLM}&\footnotesize{MSN}&\footnotesize{NLF}&\small{0.01}&\small{0.04}&\small{0.1}&\small{0.3}\\
    \midrule
    \scriptsize{\XSolidBrush} & \checkmark & \checkmark & \checkmark & 22.71 & 25.36 & 27.79& 32.27\\

    $\checkmark $ & \scriptsize{\XSolidBrush} & $\checkmark $ & $\checkmark $ & 22.78 & 25.39 & 27.86& 32.28\\

    $\checkmark $ & $\checkmark $ & \scriptsize{\XSolidBrush} & $\checkmark $ & 22.40 & 25.02 & 27.49& 31.95\\
    $\checkmark $ & $\checkmark $ & $\checkmark $ & \scriptsize{\XSolidBrush} & 22.65 & 25.25 & 27.70& 32.13\\
    \midrule
    \checkmark & $\checkmark $ &$\checkmark $ &$\checkmark $ & \textbf{22.85} &  \textbf{25.48} & \textbf{27.95}& \textbf{32.36}\\
  \bottomrule
\end{tabular}

\label{tab:7}
\end{table}


\vspace{-0.1in}
\section{EXPERIMENTAL RESULTS}

\label{section:a4}

In this section, we first elaborate the loss function, and then demonstrate the experimental settings and implementation details as well as the experimental comparisons against the existing state-of-the-art CS methods.
\vspace{-0.08in}
\subsection{Loss Function}
\vspace{-0.01in}

Given the input image $x_{i}$, the mission of the proposed NL-CSNet is to narrow down the gap between the output and the target image $x_{i}$. In addition, considering the coupling constraint between the non-local self-similarity knowledge, the total loss function can be expressed as
\vspace{-0.08in}
\begin{equation}
L = L_{r} + \gamma L_{c}
\label{eq:8}
\vspace{-0.06in}
\end{equation}
where $L_{r}$ and $L_{c}$ are the reconstruction loss and the non-local coupling loss respectively. $\gamma$ is a hyper parameter to control the non-local coupling loss item. Due to the non-local neural network is embedded into the measurement domain and multi-scale feature domain in the proposed NL-CSNet, the non-local coupling loss $L_{c}$ can be divided into two subitems: $L_{c} = \gamma_{u} L_{u} + \gamma_{v} L_{v}$, where $L_{u}$ and $L_{v}$ are the non-local coupling loss in terms of these two types of non-local models respectively. $\gamma_{u}$ and $\gamma_{v}$ are the regularization parameters to control the trade-off between these two subitems.

In order to explain the loss function more conveniently, the outputs of NL-CSNet are defined and analyzed below. Specifically, given the input image $x_{i}$, through the pipeline of the proposed entire CS framework, three entities are revealed, i.e., $r_{i}$, $\{ r_{i}^{s_{t}}\}$ and $\tilde{x}_{i}$. More specifically, $r_{i}$ indicates the produced affinity matrix of the non-local subnetwork in the measurement domain, $\{ r_{i}^{s_{t}}\}$ signifies the set of the affinity matrices for the non-local subnetwork in the multi-scale feature domain ($r_{i}^{s_{t}}$ is the affinity matrix of the $t$-th non-local submodule in the $s$-th scale space, where $s=\{1,2,...S_{B}\}$, $t=\{1,2,...S_{N}\}$). $\tilde{x}_{i}$ is the revealed reconstructed image.

For the reconstruction loss $L_{r}$, we directly use the L2 norm to constrain the distance between the reconstructed image $\tilde{x}_{i}$ and the ground truth $x_{i}$. i.e.,
\vspace{-0.05in}
\begin{equation}
L_{r} =  \frac{1}{2K} \sum_{i=1}^{K}\| \tilde{x}_{i} - x_{i}\|_{F}^{2}
\vspace{-0.04in}
\end{equation}
where $K$ indicates the batchsize of the training data.

\begin{figure}[t]

\begin{minipage}[t]{0.10\textwidth}
\centering
\includegraphics[width=0.82in]{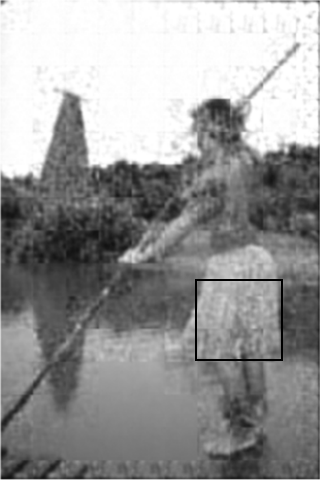}
\begin{scriptsize}

\end{scriptsize}
\end{minipage}
\hspace{0.065in}
\begin{minipage}[t]{0.10\textwidth}
\centering
\includegraphics[width=0.82in]{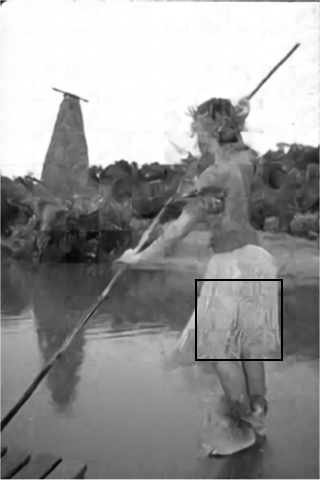}
\begin{scriptsize}

\end{scriptsize}
\end{minipage}
\hspace{0.065in}
\begin{minipage}[t]{0.10\textwidth}
\centering
\includegraphics[width=0.82in]{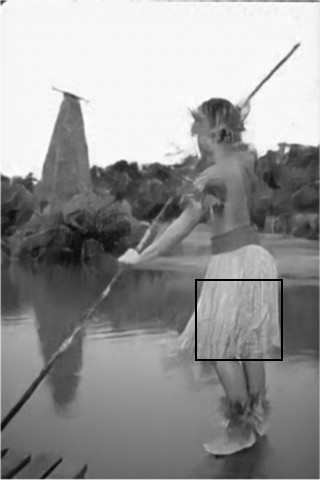}
\begin{scriptsize}

\end{scriptsize}
\end{minipage}
\hspace{0.065in}
\begin{minipage}[t]{0.10\textwidth}
\centering
\includegraphics[width=0.82in]{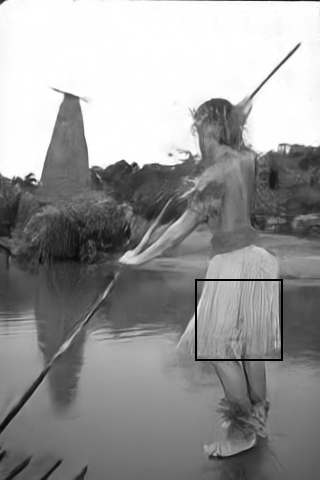}
\begin{scriptsize}

\end{scriptsize}
\end{minipage}
\vspace{-0.08in}
\\
\begin{minipage}[t]{0.10\textwidth}
\centering
\includegraphics[width=0.82in]{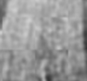}
\begin{scriptsize}

\centering
\vskip -0.52 cm \begin{tiny}ReconNet$\backslash$23.76$\backslash$0.6177\end{tiny}
\end{scriptsize}
\end{minipage}
\hspace{0.065in}
\begin{minipage}[t]{0.10\textwidth}
\centering
\includegraphics[width=0.82in]{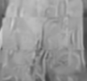}
\begin{scriptsize}

\centering
\vskip -0.52 cm \begin{tiny}ISTANet$^{\tiny{+}}$$\backslash$26.66$\backslash$0.7759\end{tiny}
\end{scriptsize}
\end{minipage}
\hspace{0.065in}
\begin{minipage}[t]{0.10\textwidth}
\centering
\includegraphics[width=0.82in]{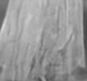}
\begin{scriptsize}

\centering
\vskip -0.52 cm \begin{tiny}\hspace{0.01in}DPA-Net$\backslash$26.25$\backslash$0.7917\end{tiny}
\end{scriptsize}
\end{minipage}
\hspace{0.065in}
\begin{minipage}[t]{0.10\textwidth}
\centering
\includegraphics[width=0.82in]{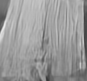}
\begin{scriptsize}

\centering
\vskip -0.52 cm \begin{tiny}NLCSNet$\backslash$\textbf{26.85}$\backslash$\textbf{0.7984}\end{tiny}
\end{scriptsize}
\end{minipage}

\vspace{-0.16in} \caption{Visual quality comparisons of deep network-based CS methods using random sampling matrix on one sample image from BSD68 in case of sampling rate = 0.1. (The ground truth image is shown in Fig.~\ref{fig:8}.)}

\label{fig:9}
\vspace{-0.22in}
\end{figure}

For the non-local coupling loss item $L_{u}$ in the measurement domain, we constrain the affinity matrix $r_{i}$ according to Eq.~\eqref{eq:6}. Since the elements in the symmetrical  positions of $r_{i}$ are the affinity coefficients between the given two measurements of referencing to each other, the loss function $L_{u}$ can be expressed as
\vspace{-0.05in}
\begin{equation}
L_{u} =  \frac{1}{2K} \sum_{i=1}^{K}\| r_{i}-{r_{i}}^{T} \|_{F}^{2}
\vspace{-0.04in}
\end{equation}
where $T$ is the transpose operator of the affinity matrix.

Analogically, for the non-local coupling loss item $L_{v}$ in the multi-scale feature domain, since multiple non-local submodules are included in the multi-scale space, the loss function $L_{v}$ can be expressed as
\vspace{-0.05in}
\begin{equation}
L_{v} =  \frac{1}{2K} \sum_{i=1}^{K}\sum_{s=1}^{S_{B}}\sum_{t=1}^{S_{N}}\| r_{i}^{s_{t}} - {r_{i}^{s_{t}}}^{T} \|_{F}^{2}
\vspace{-0.03in}
\end{equation}
where $S_{B}$ indicates the number of the scale spaces and $S_{N}$ is the number of the non-local submodules in each scale space.

\begin{figure}[b]
\begin{center}
\vspace{-0.22in}
\includegraphics[width=1.7in]{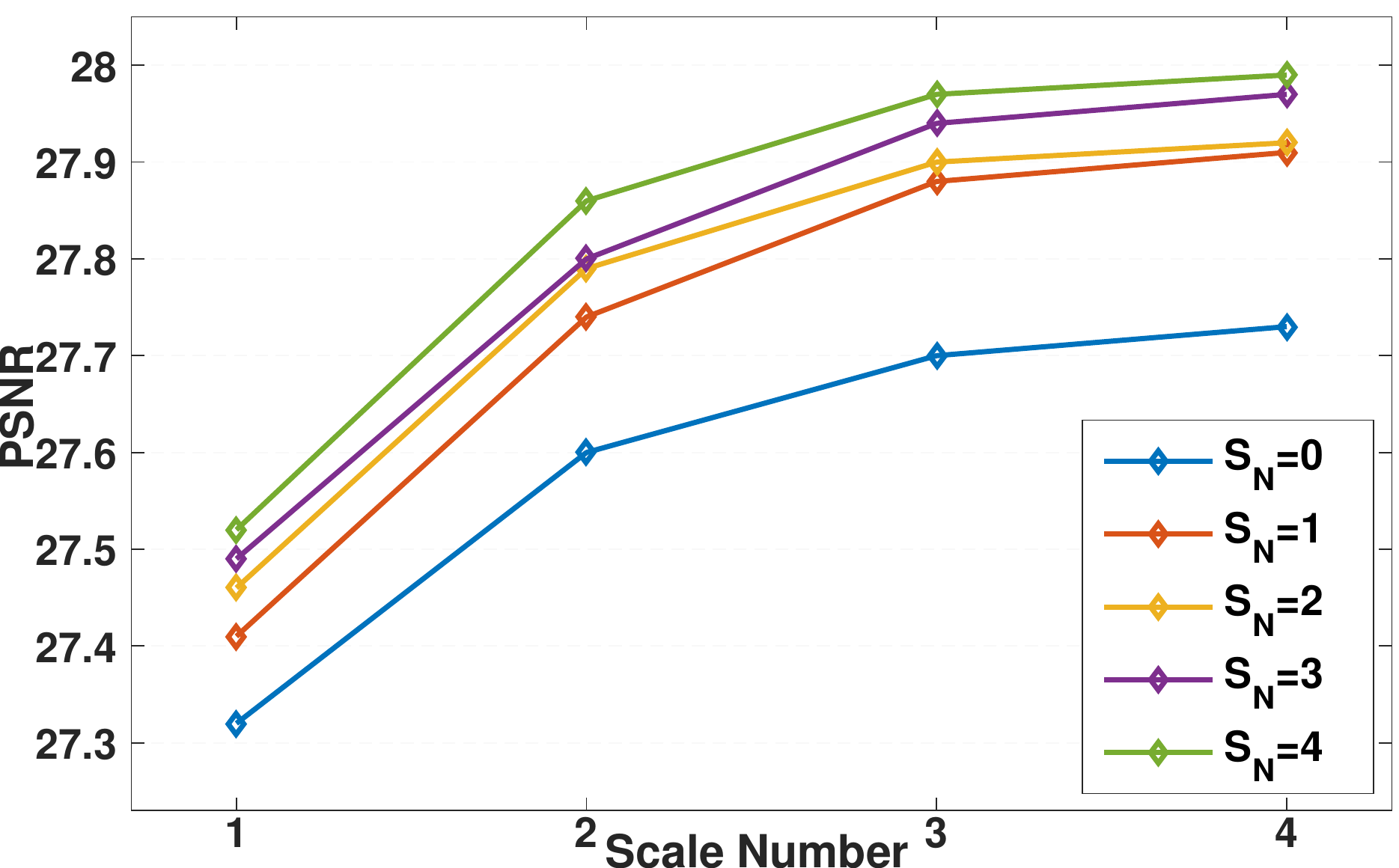}
\includegraphics[width=1.7in]{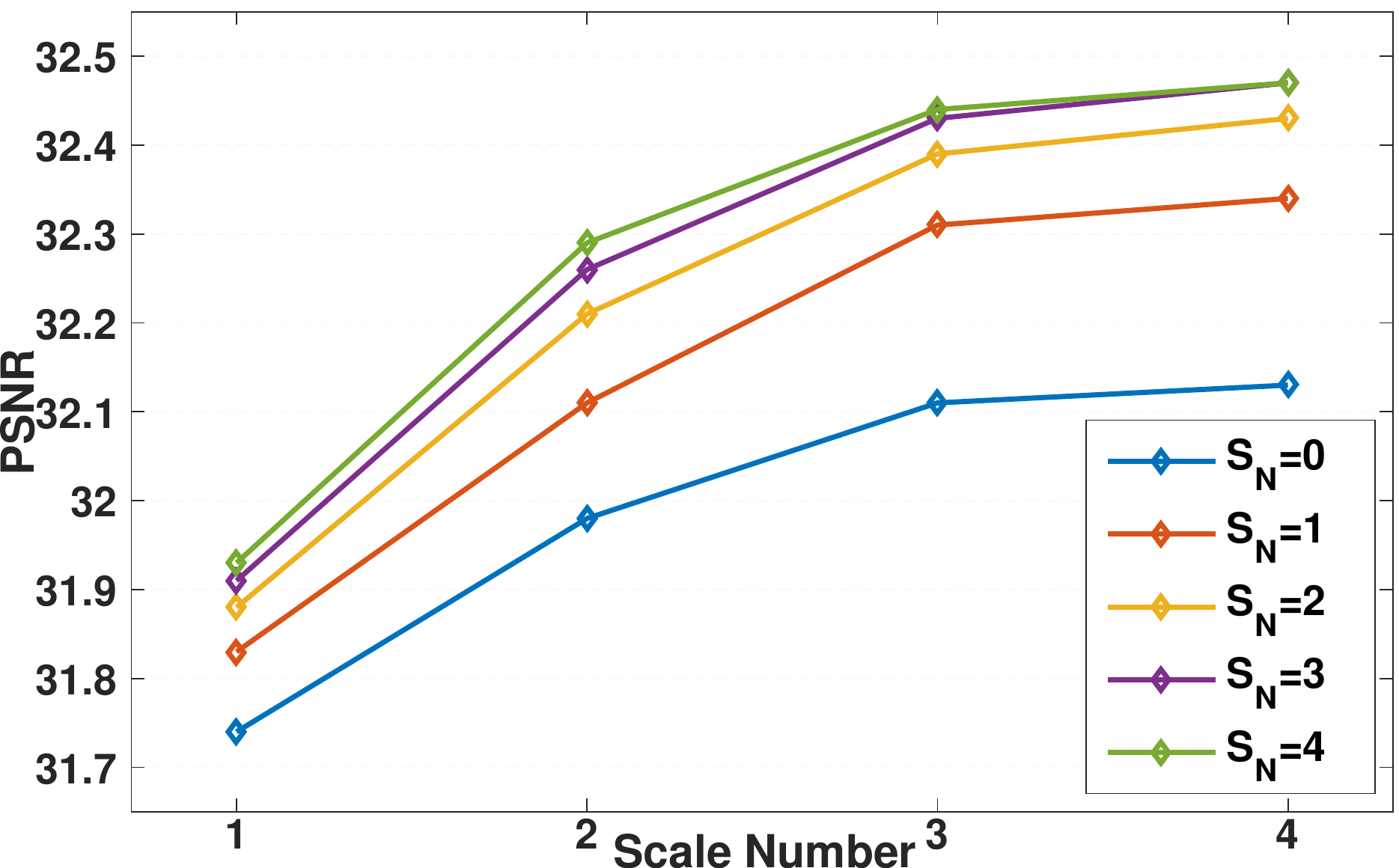}
\end{center}
\vspace{-0.09in} \scriptsize{\quad\quad\quad\quad\quad\quad sampling rate = 0.1 \quad\quad\quad\quad\quad\ \quad\quad\quad\quad sampling rate = 0.3}
\vskip -0.05in
   \caption{The relationship between the hyperparameters ($S_{B}$, $S_{N}$) and the reconstruction quality in the proposed model. The X-axis represents the scale number $S_{B}$, the Y-axis is the reconstruction quality, and the curves signify the relationship between $S_{B}$ and reconstruction quality under different $S_{N}$.}

\label{fig:10}
\end{figure}

\vspace{-0.11in}
\subsection{Implementation and Training Details}
\vspace{-0.01in}

In the proposed CS framework, we set block size $B$=32 as the previous works~\cite{ref6,ref21}. For the hyperparameters $S_{B}$ and $S_{N}$ of the deep reconstruction subnetwork MS-NLNet, Fig.~\ref{fig:10} reveals the relationship between these two hyperparameters and the reconstruction quality, from which we observe that with the increase of the hyperparameters, the quality becomes more and more insensitive to them. In our model, we set the number of the horizontal branches as 3, which implies 3 scale spaces are considered, i.e., $S_{B}=3$. In each scale space, we set the number of non-local submodules as 3, i.e., $S_{N}=3$. The kernel number of the convolutional layers (channel number $C$ of the feature maps) in the three horizontal branches are separately set as 16, 32 and 64. Besides, for the vertical branches, we set the number of both downsampling and upsampling branches as 3, i.e., $d_{u}=d_{v}=3$ (two downsampling and upsampling branches are shown in Fig.~\ref{fig:4}). In terms of the number of the residual blocks, we set $d_{l}=d_{r}=3$ in our model. In addition, the numbers of the residual blocks ($d_{t}$) in the non-local submodules of the three horizontal branches are different and we set them as 1, 2 and 3 respectively. To reduce the resource consumptions of the non-local neural network in the multi-scale feature domain, we use the same spatial downsampling trick ($\downarrow$ in Fig.~\ref{fig:4}) as~\cite{ref65}. Specifically, for the three scale spaces in our framework, we do downsampling operations at the scale factors of 16, 4 and 1 respectively. Furthermore, for the downsample block and upsample block in the vertical branches, a convolutional layer with stride 2 and a pixelshuffle layer is applied respectively for feature maps downsampling and upsampling. It is worth noting that the kernel size of all the convolutional layers are set as 3$\times$3, except the convolutional layers in the aggregation submodules and the non-local networks. For training configurations, we initialize the convolutional filters using the same method as~\cite{ref69} and pad zeros around the boundaries to keep the size of feature maps the same as the input.

\begin{figure*}[t]

\begin{minipage}[t]{0.135\textwidth}
\centering
\includegraphics[width=0.95in]{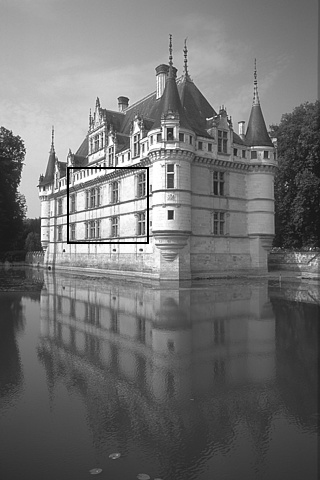}
\begin{scriptsize}

\end{scriptsize}
\end{minipage}
\hspace{-0.012in}
\begin{minipage}[t]{0.135\textwidth}
\centering
\includegraphics[width=0.95in]{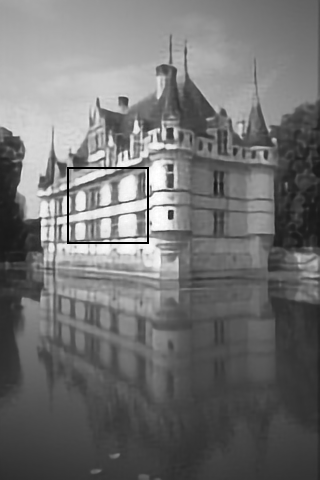}
\begin{scriptsize}

\end{scriptsize}
\end{minipage}
\hspace{-0.012in}
\begin{minipage}[t]{0.135\textwidth}
\centering
\includegraphics[width=0.95in]{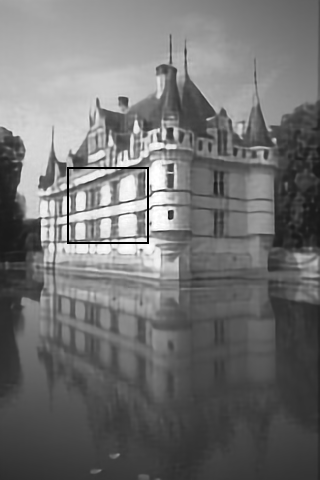}
\begin{scriptsize}

\end{scriptsize}
\end{minipage}
\hspace{-0.012in}
\begin{minipage}[t]{0.135\textwidth}
\centering
\includegraphics[width=0.95in]{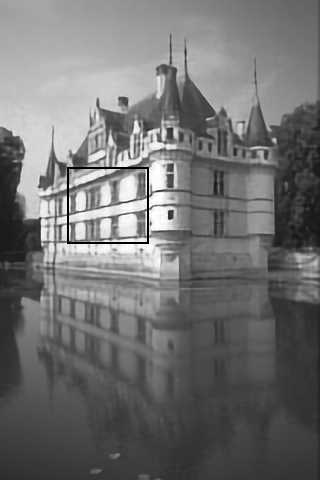}
\begin{scriptsize}

\end{scriptsize}
\end{minipage}
\hspace{-0.012in}
\begin{minipage}[t]{0.135\textwidth}
\centering
\includegraphics[width=0.95in]{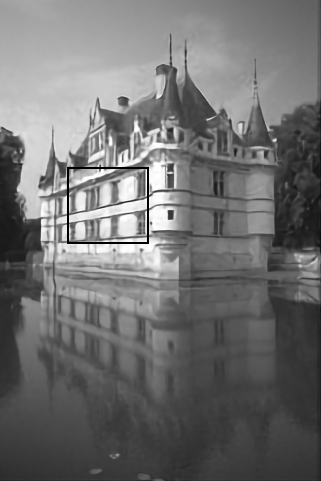}
\begin{scriptsize}

\end{scriptsize}
\end{minipage}
\hspace{-0.012in}
\begin{minipage}[t]{0.135\textwidth}
\centering
\includegraphics[width=0.95in]{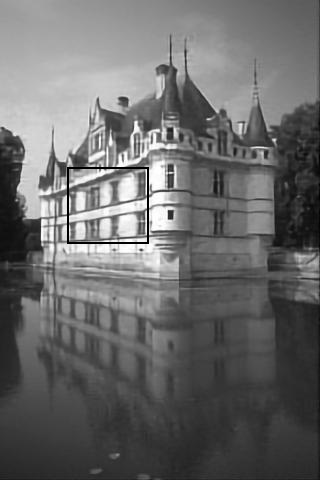}
\begin{scriptsize}

\end{scriptsize}
\end{minipage}
\hspace{-0.012in}
\begin{minipage}[t]{0.135\textwidth}
\centering
\includegraphics[width=0.95in]{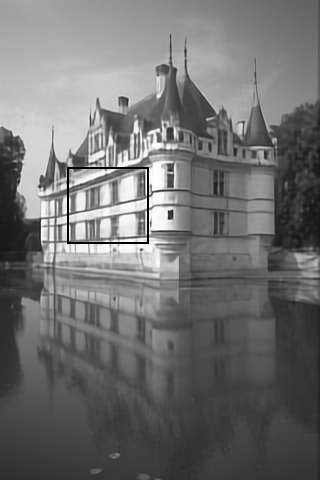}
\begin{scriptsize}

\end{scriptsize}
\end{minipage}
\vspace{-0.13in}
\\
\begin{minipage}[t]{0.135\textwidth}
\centering
\includegraphics[width=0.95in]{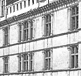}
\begin{scriptsize}
\centering
\vskip -0.52 cm \begin{tiny}Methods$\backslash$PSNR$\backslash$SSIM\end{tiny}
\end{scriptsize}
\end{minipage}
\hspace{-0.012in}
\begin{minipage}[t]{0.135\textwidth}
\centering
\includegraphics[width=0.95in]{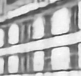}
\begin{scriptsize}
\centering
\vskip -0.52 cm \begin{tiny}CSNet~\cite{ref55}$\backslash$26.22$\backslash$0.8307\end{tiny}
\end{scriptsize}
\end{minipage}
\hspace{-0.012in}
\begin{minipage}[t]{0.135\textwidth}
\centering
\includegraphics[width=0.95in]{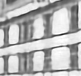}
\begin{scriptsize}
\centering
\vskip -0.52 cm \begin{tiny}CSNet$^{+}$~\cite{ref6}$\backslash$26.30$\backslash$0.8356\end{tiny}
\end{scriptsize}
\end{minipage}
\hspace{-0.012in}
\begin{minipage}[t]{0.135\textwidth}
\centering
\includegraphics[width=0.95in]{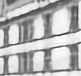}
\begin{scriptsize}
\centering
\vskip -0.52 cm \begin{tiny}SCSNet~\cite{ref19}$\backslash$26.44$\backslash$0.8378\end{tiny}
\end{scriptsize}
\end{minipage}
\hspace{-0.012in}
\begin{minipage}[t]{0.135\textwidth}
\centering
\includegraphics[width=0.95in]{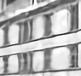}
\begin{scriptsize}
\centering
\vskip -0.52 cm \begin{tiny}OPINENet$^{\tiny{+}}$~\cite{ref23}$\backslash$26.89$\backslash$0.8544\end{tiny}
\end{scriptsize}
\end{minipage}
\hspace{-0.012in}
\begin{minipage}[t]{0.135\textwidth}
\centering
\includegraphics[width=0.95in]{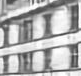}
\begin{scriptsize}
\centering
\vskip -0.52 cm \begin{tiny}AMP-Net~\cite{ref22}$\backslash$26.96$\backslash$0.8472\end{tiny}
\end{scriptsize}
\end{minipage}
\hspace{-0.012in}
\begin{minipage}[t]{0.135\textwidth}
\centering
\includegraphics[width=0.95in]{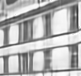}
\begin{scriptsize}
\centering
\vskip -0.52 cm \begin{tiny}NL-CSNet$^{\ast}$$\backslash$\textbf{27.12}$\backslash$\textbf{0.8686}\end{tiny}
\end{scriptsize}
\end{minipage}

\vspace{-0.16in} \caption{Visual quality comparisons of deep network-based CS methods using learned sampling matrix on one sample image from BSD68 in case of sampling rate = 0.1.}

\label{fig:11}
\vspace{-0.22in}
\end{figure*}

We use training set (400 images) from BSD500~\cite{ref70} dataset and the training set of VOC2012~\cite{ref71} as our training data. Specifically, in the training process, we use a batch size of 8 and randomly crop the size of patches to 128$\times$128. We augment the training data in two ways: ($1$) Rotate the images by 90$^{\circ}$, 180$^{\circ}$ and 270$^{\circ}$ randomly. ($2$) Flip the images horizontally with a probability of 0.5. We use the Pytorch toolbox and train our model using the Adaptive moment estimation (Adam) solver on a NVIDIA GTX 1080Ti GPU. In addition, we set $\gamma = 0.001$, $\gamma_{u} = 1.0$ and $\gamma_{v}=1.0$ in our model. We set the momentum to 0.9 and the weight decay to 1e-4. The learning rate is initialized to 1e-4 for all layers and decreased by a factor of 2 for every 30 epochs. We train our model for 200 epochs totally and 1000 iterations are performed for each epoch. Therefore 200$\times$1000 iterations are completed for the whole training process.

\vspace{-0.1in}
\subsection{Comparisons with State-of-the-art Methods}
\vspace{-0.02in}

To evaluate the performance of the proposed CS framework, we conduct the comparisons against the existing CS methods in terms of two aspects: reconstruction quality comparisons and running speed comparisons. Depending on whether the sampling matrix can be learned, the comparisons of the CS reconstruction methods are provided in two aspects: comparisons with the random sampling matrix-based CS methods and the learned sampling matrix-based CS methods. a) For the random sampling matrix-based CS methods, the optimization-based and the deep network-based reconstruction methods are both concerned. Specifically, considering the optimization-based methods, three representative CS schemes are selected, i.e., TV~\cite{ref53}, MH~\cite{ref32} and GSR~\cite{ref31}. In view of the deep network-based methods, more than ten CS algorithms are considered, including: SDA~\cite{ref33}, ReconNet~\cite{ref16}, LDIT~\cite{ref20}, LDAMP~\cite{ref20}, DPDNN~\cite{ref63}, I-Recon~\cite{ref17}, ISTA-Net~\cite{ref21}, DR$^{2}$-Net~\cite{ref15}, IRCNN~\cite{ref42}, NN~\cite{ref64}, NLR-CSNet~\cite{ref54}, DPA-Net~\cite{ref37} and DPIR~\cite{ref44}. b) For the learned sampling matrix-based CS methods, seven recent literatures, i.e., CSNet~\cite{ref55}, LapCSNet~\cite{ref35}, SCSNet~\cite{ref19}, CSNet$^{+}$~\cite{ref6}, BCS-Net~\cite{ref39}, OPINE-Net$^{+}$~\cite{ref23} and AMP-Net~\cite{ref22} participate in the comparison in our experiments.

For the fairness of comparison, two model variants, dubbed NL-CSNet and {NL-CSNet}$^{\ast}$, are generated in our experiments by using random sampling matrix and learned sampling matrix. For NL-CSNet, the orthogonalized Gaussian random matrix~\cite{ref37,ref54} is utilized in our experiments, and during the training process, the sampling matrix remains unchanged. For {NL-CSNet}$^{\ast}$, the sampling matrix is optimized jointly with the reconstruction network. For testing data, we carry out extensive experiments on several representative benchmark datasets: Set5~\cite{ref19}, Set14~\cite{ref35}, Set11~\cite{ref72} and BSD68~\cite{ref37}, which are widely used in the previous CS-related works. To ensure the fairness of the comparison, we evaluate the reconstruction performance with two widely used quality evaluation metrics: PSNR and SSIM in terms of various sampling rates.

\subsubsection{Reconstruction Quality Comparisons}

In our experiments, the compared CS methods are divided into the following two groups: random sampling matrix-based CS methods and learned sampling matrix-based CS methods.

\textbf{Random sampling matrix-based CS methods:} For the random sampling matrix-based CS methods, two types of CS algorithms, i.e., the optimization-based methods and the deep network-based methods are mainly concerned. \textbf{a)} For the optimization-based methods, since the performance of the method GSR is better than the other two algorithms TV and MH, we mainly analyze the experimental results compared with GSR in our experiments. The quantitative results on dataset Set5 are shown in Table~\ref{tab:1}, from which we can get that the proposed NL-CSNet outperforms other optimization-based CS methods. Specifically, for the given five sampling rates from 0.01 to 0.30 (i.e., 0.01, 0.04, 0.10, 0.20 and 0.30), the proposed framework achieves on average 0.87dB and 0.0331 gains in PSNR and SSIM compared against GSR on the dataset Set5. Especially at low sampling rate (e.g., 0.01), the proposed method can obtain more than 3dB gain compared against GSR. The visual comparisons are displayed in Fig.~\ref{fig:5}, from which we observe that the proposed NL-CSCNet is capable of preserving more structural details compared with the other optimization-based CS methods. \textbf{b)} For the deep network-based CS methods, Tables~\ref{tab:1} and~\ref{tab:3} respectively present the experimental comparisons in terms of the given sampling rates on two datasets Set5 and Set11, from which we can observe that the proposed NL-CSNet achieves competitive or even superior performance against the existing deep network-based CS schemes. The visual comparisons are shown in Fig.~\ref{fig:9}, from which we observe that the proposed NL-CSCNet is capable of preserving more texture information and recovering richer structural details compared against the other deep network-based CS methods that use random sampling matrix.

\textbf{Learned sampling matrix-based CS methods:} For the learned sampling matrix-based CS methods, the sampling matrix is optimized jointly with the reconstruction module, which facilitates the collaborations between the sampling and reconstruction. For comparative fairness, in our experimental variant NL-CSNet$^{*}$, the sampling matrix is also optimized jointly with the reconstruction process. Tables~\ref{tab:2},~\ref{tab:4} and~\ref{tab:5} separately present the experimental comparisons in terms of the given five sampling rates (i.e., 0.01, 0.04, 0.10, 0.20 and 0.30) on three datasets (Set5, Set11 and Set14), from which we can get that the proposed NL-CSNet$^{*}$ performs much better than the other deep network-based CS schemes. In the compared CS methods, the recent algorithms SCSNet~\cite{ref19}, CSNet$^{+}$~\cite{ref6} and AMP-Net~\cite{ref22} can obtain the best reconstruction performance. For simplicity, we mainly analyze the experimental results compared with these three representative CS algorithms. Specifically, (1) On the dataset Set5, the proposed framework achieves on average 0.74dB, 0.87dB, 0.63dB and 0.0188, 0.0203, 0.0143 gains in PSNR and SSIM compared against these three deep network-based CS algorithms in terms of the given sampling rates. (2) On the dataset Set11, the proposed framework achieves on average 1.20dB, 1.34dB, 0.66dB and 0.0325, 0.0343, 0.0216 gains in PSNR and SSIM compared against the three CS methods in terms of different sampling rates. (3) On the dataset Set14, the proposed framework achieves on average 0.73dB, 0.83dB, 0.34dB and 0.0225, 0.0234, 0.0190 gains in PSNR and SSIM in terms of various sampling rates. The visual comparisons are shown in Figs.~\ref{fig:8} and~\ref{fig:11}, from which we observe that the proposed method is capable of preserving more details and retaining sharper edges compared to the other representative deep network-based CS methods.

\subsubsection{Running Speed Comparisons}
\vspace{-0.01in}

To verify the efficiency of the proposed CS framework, we also display the reconstruction speed of different CS methods. Specifically, we evaluate the runtime on the same platform with 3.30 GHz Intel i7 CPU (32G RAM) plus NVIDIA GRX 1080Ti GPU (11G Memory). Table~\ref{tab:6} shows the average running time comparisons (in second) between different CS methods (including the optimization-based and deep network-based CS methods) for reconstructing a 256$\times$256 image at two sampling rates of 0.01 and 0.10. It is worth noting that the running time result of the algorithm SDA is copied from~\cite{ref16}, and for the other CS methods, we test them on the same platform with their source codes downloaded from the authors' websites. In addition, the optimization-based CS schemes are implemented based on CPU device. In contrast, we test all the deep network-based CS methods on both the CPU and GPU devices. The running speed comparison results show that the deep network-based methods run faster than the optimization-based methods. Furthermore, the proposed NL-CSNet basically remains at the same order of magnitude as the other existing deep network-based methods and achieves a more faster reconstruction compared to the other optimization-based CS algorithms.

\vspace{-0.1in}
\subsection{Ablation Studies and Discussions}
\vspace{-0.02in}

As noted above, the proposed CS framework achieves higher reconstruction quality. In this subsection, we mainly analyze the contribution of each submodule of the proposed CS framework. For image sampling, as shown in Tables~\ref{tab:1} and~\ref{tab:2} as well as Tables~\ref{tab:3} and~\ref{tab:4}, the learned sampling matrix obtains more than 2dB gain on average compared to the Gaussian random sampling matrix. For image reconstruction, in order to evaluate the benefits of each part of the proposed CS reconstruction network, we design several counterpart versions of the proposed reconstruction model, in which some functional modules are selectively discarded or retained. Table~\ref{tab:7} shows the experimental results on the dataset BSD68~\cite{ref37}, in which four functional modules, i.e., non-local coupling loss item (Coupling), non-local in the measurement domain (NLM), multi-scale network architecture (MSN) and non-local in the multi-scale feature domain (NLF) are considered. Specifically, the check marks and the cross marks indicate the reserving and discarding of the corresponding functional modules respectively. It should be noted that when MSN is discarded, we only use the first horizontal branch of MS-NLNet to reconstruct the target image. The experimental results in Table~\ref{tab:7} reveal that the multi-scale network architecture (MSN) brings a maximum gain, and the non-local in the measurement domain (NLM) brings a minimum gain. Besides, the non-local in the multi-scale feature domain (NLF) can bring more growth against that of the measurement domain (NLM).

In view of the non-local coupling loss item (Coupling) of the proposed model, its mission is to enhance the coupling between the non-local self-similarity knowledge. In order to choose a appropriate regularization parameter $\gamma$ in Eq.~\eqref{eq:8}, we conduct a large number experimental comparisons in case of different settings of $\gamma$. Considering $\gamma=0$ as a baseline (same with the traditional non-local neural networks~\cite{ref65}), we observe that with the increase of $\gamma$, the reconstructed quality can be improved to a certain extent, but when $\gamma$ is too large, the reconstruction will be corroded slightly. Moreover, the visual comparisons of the affinity matrices produced in the non-local subnetwork of measurement domain in terms of different $\gamma$ are shown in Fig.~\ref{fig:7}, from which we can get that when $\gamma=0$, the localization of similar patches is inefficient, and when $\gamma$ is too large, the learned affinities are tend to 0. Through experimental analysis, we finally set $\gamma$ as 0.001. It is worth noting that when the non-local coupling loss item (Coupling) is discarded, it actually is the traditional non-local neural network~\cite{ref65}. Table~\ref{tab:7} shows that the proposed non-local coupling loss item is capable of improving the CS reconstruction performance to a certain extent. As mentioned above, each of the aforementioned four functional modules in the proposed NL-CSNet can enhance the CS reconstruction quality to varying degrees.


\vspace{-0.08in}
\section{Conclusion}
\label{section:a6}

In this paper, we propose a novel image compressed sensing framework using non-local neural network, which utilizes the non-local self-similarity priors with deep network to improve the reconstruction quality. In the proposed model, two non-local subnetworks are designed to establish the long-distance references between the non-local self-similarity knowledge in the measurement domain and multi-scale feature domain respectively. Specifically, in the subnetwork of measurement domain, the affinities between the measurements of different image blocks are explored for mining the interblock correlations. Analogically, in the subnetwork of multi-scale feature domain, the dependencies between the non-local feature representations are explored for building a reference mechanism between the non-local structural textures. Experimental results demonstrate that the proposed CS framework achieves much higher reconstruction performance and better perceptual image quality against other state-of-the-art CS methods.




\vfill

\end{document}